\documentclass[10pt,twocolumn,article]{memoir} 
\pdfoutput=1
\usepackage[T2A]{fontenc}
\usepackage[utf8]{inputenc}
\usepackage{graphicx}
\usepackage{amsmath}
\usepackage{amsthm}
\usepackage{amsfonts}
\usepackage{latexsym, amssymb}
\usepackage{hyperref}
\usepackage{appendix}
\usepackage[text={7in,8.5in},centering]{geometry}
\usepackage{microtype} 
\makeatletter
\def\MT@register@subst@font{\MT@exp@one@n\MT@in@clist\font@name\MT@font@list
\ifMT@inlist@\else\xdef\MT@font@list{\MT@font@list\font@name,}\fi}
\makeatother
\captiontitlefont{\small\itshape} 
\captionnamefont{\small\itshape} 
\counterwithout{section}{chapter}
\counterwithin{subsection}{section}
\begin{document}

\title{Special relativity and the twins: a review and a new approach}
\author{David Derbes\thanks{\url{derbes.physics@gmail.com} }\\
\small \emph{1700 E.\,56th St., Apt.\,2007}\\ \small \emph{Chicago, IL 60637}\normalsize}

\maketitle

\begin{abstract}
It is sometimes claimed that the twin ``paradox'' requires general relativity for a resolution. This paper presents a simple, exact resolution using only special relativity and the equivalence principle. Two earlier approximate solutions are considered, along with some background review to render the article self-contained. It is hoped that this material will be suitable for classroom instruction.
\vspace*{0.1in}
\end{abstract}

\setcounter{section}{-1}

\section{Prolegomena and apologia}

Full disclosure: Shorter versions of this article have been rejected four times over thirty-five years. The objections (after the first, in 1988) were broadly two: that there was nothing new in it, but more forcefully, that articles on the twins were harmful. Not in themselves, but in the second order: these articles frequently induce a cloud of cranks who swamp journals in the misbegotten hope of disproving relativity and Einstein, thereby obliging conscientious editors to waste valuable time and energy refuting nonsense. So why write another article? Why read one? 

The selfish motive is to publish what this author thinks, \emph{pace} his referees, is a new approach to this old puzzle.\footnote{Careful review of the literature years ago failed to find an \emph{exact} special relativistic solution in the literature. The author recently discovered the paper by J.\,Gamboa, F.\,Mendez, M.\,B.\,Paranjape and Benoit Sirois, ``The twin paradox: the role of acceleration'', \emph{Can.\,Jour.\,Phys.}\,\textbf{97} (2019) 1049, arXiv:gr-qc/1807.02148v1, which \emph{is} exact and special relativistic. While the results and some of the calculations are very similar to those of this paper, the approach is very different.} Less selfish motives are pedagogic, to clear up a number of misunderstandings related to the twins. Some hold that special relativity can provide only an approximate resolution to the puzzle; exact reconciliation of the twins' times requires the general theory. Occasionally one reads that special relativity applies only to inertial reference frames and cannot handle accelerated frames. As will be shown, each of these claims is mistaken. Additionally, many students have at best a murky understanding of how acceleration affects clock rates, which this paper may clear up. The author also hopes to make better known the use of Møller coordinates and the encyclopedic writings of H.\,Arzèlies. Perhaps this elementary paper will provide something useful to those who teach relativity.

\section{Introduction}

On January 16, 1911, Einstein gave a lecture to the Zurich Society of Natural Sciences, entitled \emph{The Theory of Relativity}.\footnote{A.\,Einstein, ``Die Relativit\"{a}tstheorie,'' \emph{Vierteljahrsschrift} \textbf{56} 1--14 (1911). English translation in \emph{The Collected Papers of Albert Einstein, Vol.\,3: The Swiss Years: Writings 1909--1911}, (Princeton U.\,P., 1994), document 17, 340--350. This quote is from pp.\,348--349. Einstein's papers are freely available online in the original language and in English: \url{https://einsteinpapers.press.princeton.edu/} .} After discussing the usual time dilation, he added that it was ``at its strangest'' if one imagined a clock given a uniform velocity in one direction for some distance, then returning with the reverse velocity, coming to rest at its original position. He went on:
\begin{quote}\small
Were we, for example, to place a living organism in a box and make it perform the same to-and-fro motion as the clock discussed above, it would be possible to have this organism return to its original starting point after an arbitrarily long flight having undergone an arbitrarily small change, while identically constituted organisms that remained at rest at the point of origin have long since given way to new generations. \end{quote}\normalsize
In the same year, Einstein's friend Paul Langevin published an article\footnote{P.\,Langevin, ``L'evolution de l'espace et du temps,'' \emph{Scientia}, \textbf{10} 31--54 (1911). This article was republished in English: ``The evolution of space and time,'' \emph{Scientia}, \textbf{108} 285--300 (1973): \url{https://amshistorica.unibo.it/7} .} in which the living organism was a space traveler. At a speed ``sufficiently close to that of light'' that would allow him to reach a star in a time of one year, upon his return he would find that two centuries had passed on Earth. The dramatic return of the \emph{voyageur de Langevin} became all the more forceful when the traveler was identified as one of two twins, the other remaining behind.\footnote{Previously the problem was called ``the clock paradox''. The earliest reference to twins may be in R.\,M.\,Frye and V.\,M.\,Brigham, ``Paradox of the Twins'', \emph{Am.\,J.\,Phys.}\,\textbf{25}, 553-555 (1957).} Various authors\footnote{For a survey of the literature prior to 1969, see L.\,Marder, \emph{Time and the Space-Traveller}, (U.\ Pennsylvania Press, Philadelphia, 1971).\label{fn:Marder1971}} (including Einstein himself\footnote{``Dialog über Einwände gegen die Relativitätstheorie'' (Dialog concerning objections to the theory of relativity), \emph{Naturwis.}\,\textbf{6} (1918). Einstein outlined but did not present the calculation given by Born (note \ref{fn:Born}) two years later (see \S4, below).\label{fn:Einstein_dialog}}) considered the traveler over the next several years.

In the late 1950's, a raging controversy on the twins played out in various journals, with those of opposing viewpoints nearly coming to (epistolary) blows. A major point of contention was the role of general relativity; was it needed or not?\footnote{For a more recent general relativistic treatment, see R.\,Perrin, ``Twin paradox: A complete treatment from the point of view of each twin,'' \emph{Am.\,J.\,Phys.}\,\textbf{47}, (1979) 317--319. Perrin uses much the same travel plan as will be considered here, but divides the distance differently. With that proviso, his paper and this one find the same values of the stay-at-home twin's time from each twin's point of view.} A reader of even a small sample of the literature will be dismayed by the variety and complexity of mathematics brought to bear by so many authors. The poet Wallace Stevens described thirteen ways of looking at a blackbird;\footnote{Wallace Stevens, ``Thirteen ways of looking at a blackbird'', in \emph{Harmonium, poems by Wallace Stevens}, (Knopf, New York, 1923), p.\,135. By coincidence Stevens and Einstein were both born in 1879 and both died in 1955.} there must be at least thirteen ways of looking at the twins. This article considers only four, the first three devised long ago. As shown here, there's no need to invoke general relativity to resolve the supposed inconsistency. The fourth, presenting an \emph{exact}, \emph{special} relativistic resolution, is new. It is explicitly dynamic (rather than kinematic), and surprisingly simple mathematically. A result from the calculus of variations is invoked, but other than that, only linear algebra and standard integrals are used. This version should be accessible to any undergraduate who has seen special relativity and Lagrangian mechanics. Readers familiar with the literature and interested only in the new resolution should skip to \S5. For \emph{lagniappe}, a puzzle of Feynman's finds an exact solution by the same methods used to resolve the twins' times. Finally, an appendix discusses the details of three graphs comparing the two twins' reckonings.

\section{The Naive View}

Let Alice and Bill be fraternal twins in a thought experiment where spaceships can travel at appreciable fractions of the speed of light. Alice blasts off for Proxima Centauri (her closest approach is exactly 4 light-years (ly) away in the Earth-Proxima frame) traveling at 80\% of the speed of light. She turns around the instant she reaches her farthest point, and comes home at the same speed. Bill stays on Earth. He calculates the time for Alice's trip as
\begin{equation}
\Delta t_{B} = \dfrac{2L}{v} = \dfrac{8\,\text{ly}}{0.8\,c} = 10\,\text{years}
\end{equation}
On the other hand, because Bill has studied special relativity, he presumes that Alice's clock runs slowly in the ratio of $1 : \gamma$, where
\begin{equation}
\gamma = \dfrac{1}{\sqrt{1 - (v/c)^{2}}} = \tfrac{5}{3}
\end{equation}
and so he believes that Alice's clock should read
\begin{equation}
\Delta t_{A}' = \dfrac{1}{\gamma}\Delta t_{B} = \bigl(\tfrac{3}{5}\bigr)\cdot 10\,\text{years} = 6\,\text{years}
\end{equation}
\indent How does Alice calculate the times of herself and her brother? She regards herself as at rest, while the Bill-Proxima reference frame first moves one way, and then the other. Because of Lorentz contraction, she measures the distance between Earth and Proxima shorter in her frame than Bill does in his:
\begin{equation}
L' = L/\gamma = \bigl(\tfrac{3}{5}\bigr)\cdot 4\,\text{ly} = 2.4\,\text{ly}
\end{equation}
In her frame Bill and the star move past her at the very same speed of 0.8 $c$. She measures a time of 
\begin{equation}
\Delta t_{A}' = \dfrac{2.4\,\text{ly}}{0.8\,c} = 3\,\text{years}
\end{equation}
for Proxima to come to her, and by symmetry an equal time for Bill to come back to her; a total time of 
\begin{equation}
\Delta t_{A}' = 6\,\text{years}
\end{equation}
exactly what Bill ascribes to her. So far, no paradox.

The difficulty arises when Alice reasons as follows: ``Since I am stationary, and Bill is moving, the time $\Delta t_{B}'$ that I believe Bill will measure on his own clock should be
\begin{equation}
\Delta t_{B}' \stackrel{?}{=} \Delta t_{A}'/\gamma = \bigl(\tfrac{3}{5}\bigr)\cdot 6\,\text{years} = 3.6\,\text{years}
\end{equation}
and so his clock should read 3.6 years when we are back together again on Earth, while my clock reads 6 years.'' Bill says, ``You're right about the 6 years on your own clock, but my clock reads 10 years, because you were moving and I was not.'' Alice replies, ``Is this a theory of relative motion, or not? As far as I'm concerned, \emph{I} was stationary, \emph{you} were moving, and so \emph{your} clock should be slow relative to mine!''

That's the ``paradox''. The naive answer is that special relativity deals only with uniform velocity; at the turnaround point, there has to be acceleration in order to reverse Alice's motion. Not only does the acceleration break the symmetry between the twins, but the non-constant speed means that special relativity does not apply, and that gets us out of the ``paradox''. Some who hold this opinion say that only in \emph{general} relativity can the apparent inconsistency be resolved.\footnote{In his \emph{Essential Relativity}, rev.\,2nd ed., (Springer, New York, 1977), W.\,Rindler writes: ``The superstition is that SR does not apply to phenomena involving accelerations: this has caused some authors to assert that GR is needed to resolve the clock paradox. In fact, SR applies to \emph{all} physics in inertial frames, and GR simply reduces to SR in such frames,'' (p.\,47).\label{fn:Rindler1977}} This is not true: special relativity \emph{can} be applied to accelerated objects and even accelerated reference frames---as will be done here---moment by moment. Below are three resolutions of the twins, of increasing complexity. None uses general relativity.

\section{Solution by Abrupt Change of Inertial Frames}

The twins call to mind a similar ``paradox'' involving clocks and a train. Let Alice be traveling in a train at a speed $v$ past her twin Bill, at rest in the station, along the common $x$--$x'$ axis. If the speed of Alice's train is $0.6\,c$, then $\gamma = \tfrac{5}{4}$, and an interval Bill measures as 10 microseconds will be measured as 8 $\mu$s by Alice. But now Alice says, ``Wait. My clock is just as good as Bill's, and my coordinate system is, too. So why can't I say Bill's clock runs slowly with respect to mine?'' And that must be so, in a theory of \emph{relativity}. The resolution comes from the failure of simultaneity.

Introduce a third observer, Charlie, at rest with respect to Bill and a distance of $L = 0.6\,c\,\times\,10\,\mu\text{s} = 1800$ m to his right. Synchronize the three clocks in the usual way, with a light flash from a source fixed midway between Bill and Charlie, and timed so that the flash strikes Alice at the moment she is alongside Bill. Let Event 1 be the point in spacetime when and where the light strikes Bill and Alice, and Event 2 be when and where Alice passes Charlie. Though Bill and Charlie believe their clocks are synchronized with Alice's, she does not agree. From the Lorentz transformations it follows that clocks separated by a distance $L$ and synchronized in their rest frame will not be synchronized in any other frame in uniform motion relative to this one. Calling $t_{B}$ and $t_{C}$ the times corresponding to Alice's time $t_{A}' = 0$, and using $x_{B} = 0$, $x_{C} = L$, at Event 1, 
\begin{equation}
t_{A}' = \gamma\bigl(t_{B} - \dfrac{vx_{B}}{c^{2}}\bigr) = 0 = \gamma(t_{B} - 0) \;\Rightarrow\; t_{B} = 0
\end{equation}
As expected, the time Alice measures on Bill's clock is 0. The time $t_{C}$ Alice observes on Charlie's clock is found in the same way:
\begin{equation}
\begin{aligned}
t_{A}' &= \gamma\bigl(t_{C} - \dfrac{vx_{C}}{c^{2}}\bigr) = 0 \\
&= \gamma(t_{C} - \dfrac{vL}{c^{2}}\bigr) \;\Rightarrow \; t_{C} = \dfrac{vL}{c^{2}}
\end{aligned}
\end{equation}
According to Alice, Charlie's clock is ahead of Bill's by a quantity $vL/c^{2} = 3.6\,\mu$s. Equivalently, one can examine the synchronization from Alice's point of view,\footnote{T.\,M.\,Helliwell, \emph{Special Relativity}, (University Books, Sausalito, 2010), pp.\,63--71. In an earlier book, \emph{Introduction to Special Relativity}, (Allyn \& Bacon, Boston, 1966), Helliwell suggests general relativity is needed to resolve the twins fully: ``As stated before, a \underline{thorough} investigation of accelerating frames of reference is both necessary and worthwhile, and leads into the fascinating subject of general relativity,'' p.\,171.\label{fn:Helliwell}} and show that in her frame of reference, the light flash hits Charlie ahead of Bill and Alice, so that Charlie's clock is advanced by exactly the same amount, $vL/c^{2}$. At Event 2, Alice is opposite Charlie and they must agree as to what is on each other's clock: $8\,\mu$s on Alice's, $10\,\mu$s on Charlie's. But on Bill's, Alice measures a time of $3.6\,\mu$s less, or $6.4\,\mu$s. Now Alice can justifiably assert that it is the clocks of Bill and Charlie running slowly with respect to her clock, and in the right ratio:
\begin{equation}
\begin{aligned}
&\left.\begin{matrix} \Delta t_{B} =  t_{2,\,B} - t_{1,\,B} = 6.4\,\mu\text{s} - 0\,\mu\text{s}\\
                     \Delta t_{C} = t_{2,\,C} - t_{1,\,C} = 10\,\mu\text{s} - 3.6\,\mu\text{s}
\end{matrix}\right\} = 6.4\,\mu\text{s}\\
 &\hphantom{0.2in}=  \bigl(\tfrac{4}{5}\bigr)\cdot 8\,\mu\text{s} = \dfrac{1}{\gamma}\Delta t_{A}'
 \end{aligned}
\end{equation}             
Alice thinks Bill and Charlie's interval between events is $1/\gamma$ of hers, and Bill and Charlie think Alice's interval is $1/\gamma$ of theirs; there is complete symmetry, as required.

Using this result it's easy to reconcile the twins' times, as is done in Helliwell (see note \ref{fn:Helliwell}). The stay at home twin, Bill, agrees with Alice about her clock reading (it is just her proper time, and all observers agree to its value). The hard part is to reconcile Alice's clock with Bill's clock when she is again on Earth. On the outward trip, it's much the same as the train example, with Charlie being replaced by a clock near Proxima, at the turnaround point (call this the ``Proxima clock''). When Alice leaves Earth, she observes the Proxima clock (or would, if she could achieve instantaneous knowledge of what it showed) reading a time of 
\begin{equation}
t_{P} = \dfrac{vL}{c^{2}} = \dfrac{(0.8\,c)(4 c\text{-yr})}{c^{2}} = 3.2\,\text{yr}
\end{equation}
When Alice arrives at the Proxima clock, her clock reads 3 years. She observes on the Proxima clock a time of 5 years (she has to, because in the Earth-Proxima system, the Proxima and Earth clocks are synchronized). According to Alice, the Earth clock reads 1.8 years when she reaches the turnaround point ($\tfrac{3}{5}$ of 3 years). That's fine with Alice, because the interval she observes on Bill's clock and on the Proxima clock are both $1/\gamma$ times the interval she measures on her own clock:
\begin{equation}
\begin{aligned}
&\left.\begin{matrix}\Delta t_{P} =  t_{2,\,P} - t_{1,\,P} = 5\,\text{yr} - 3.2\,\text{yr}\\
                     \Delta t_{B} = t_{2,\,B} - t_{1,\,B} = 1.8\,\text{yr} - 0\,\text{yr}
\end{matrix}\right\} = 1.8\,\text{yr}\\
 &\hphantom{0.2in}= \left(\tfrac{3}{5}\right)3\,\text{yr} = \dfrac{1}{\gamma}\Delta t_{A}'
 \end{aligned}
\end{equation}
so all is well. Now she has to return.

Assume that at her farthest distance from Earth she is able to reverse her velocity instantly. This requires an abrupt change into a \emph{different} coordinate system, one moving at the same uniform speed, but in the opposite direction. Presumably the return frame was synchronized in the same manner as the outward bound frame had been earlier. Now, the ``far clock'' is Earth's, and so \emph{it} will be advanced by the very same $vL/c^{2}$, 3.2 years, ahead of the Proxima clock. The Earth clock in this new frame will read the same time as the Proxima clock, 5 years, plus the advance, 3.2 years, or 8.2 years.  If an image of the Earth clock could be transmitted instantly to her as she instantaneously reverses motion, Alice would see the Earth clock jump from 1.8 to 8.2 years! Alice records an interval of 3 years for her travel back to Earth (an interval that Bill agrees with), and measures an interval for Bill of 1.8 years. But adding 1.8 years to 8.2 years gives a final time on Bill's clock of 10 years. When Alice and Bill are again face to face, they both agree that Alice's clock reads 6 years, and Bill's clock reads 10 years, and the clocks are reconciled.

This resolution, strictly kinematic and based entirely in special relativity, involves only a momentary period of (effectively) infinite acceleration, in jumping from one inertial frame to another. Rockets do not accelerate at an infinite rate, but it shows that in principle there's no need to invoke general relativity.\footnote{See, for example, R.\,A.\,Muller, ``The Twin Paradox in Special Relativity'', \emph{Am.\,J.\,Phys.} \textbf{40}, 966--969 (1972); R.\,H.\,Romer, ``Twin Paradox in Special Relativity'', \emph{Am.\,J.\,Phys.} \textbf{27}, 131--135 (1959).} 

\section{Solution with the Equivalence Principle}

Every resolution of the twins' clocks depends in some way on the acceleration. If the acceleration does nothing else, it breaks the symmetry between the traveler and the stay at home twin. So it makes sense that to consider how acceleration affects clocks. Experiment suggests that \emph{local} acceleration does not affect an \emph{ideal} clock at all. But if two clocks are separated by a distance $L$, and one is accelerated, the clocks will keep time at different rates. This follows from the equivalence principle.

Consider a set of three clocks: one at rest, a second traveling at a constant speed $v$ along the $x$ axis as it passes the first, and a third moving parallel to the $x$ axis and accelerating in that direction, attaining a speed $v$ momentarily when it passes the first. Early in the development of relativity, it was assumed that the constantly accelerating clock and the uniformly moving clock would be observed to have equal rates at the moment they were alongside each other, provided that they had at that moment the same velocity. This is called \emph{the clock hypothesis}: acceleration has no effect on the rate of an ideal clock compared to an identical, unaccelerated ideal clock at the same location traveling at the same speed. It has been established experimentally to a very high precision.\footnote{If the decay rates of muons moving in storage rings at uniform speed near $c$ depended on their accelerations, the Particle Data Group tables would so indicate; they do not. Rindler (note \ref{fn:Rindler1977}) cites (p.\,57) experiments with the ``thermal'' Doppler effect: ``One by-product of this experiment was an impressive validation of the clock hypothesis: in spite of accelerations up to $10^{16} g$(!), these nuclear `clocks' were slowed simply by the velocity factor $(1 - v^{2}/c^{2})^{1/2}$.'' (See Marder, (note \ref{fn:Marder1971}), p.\,158.) In the journey to be considered, Alice's acceleration is arranged to be even less than Earth's $g$, and it is difficult to imagine this would affect her clock in any way. A \emph{mechanical} clock could of course be accelerated (e.g., by Aaron Judge with his baseball bat) by a force sufficient for its destruction, but that clock would not qualify as ideal.} 

What if the accelerated clock is some distance away? According to the equivalence principle, there is locally no difference between a gravitational field and an accelerated reference frame. Clocks at different gravitational potential keep time at different rates. The direct proof of that is the gravitational red shift. It may be helpful to go through a quick derivation.\footnote{What follows is \emph{not} the method Einstein used in 1911; he used the Doppler shift and avoided $E = hf$. It is a variation on a method employed in that same paper: A.\,Einstein, ``\"{U}ber den Einflu{\ss} der Schwerkraft auf die Ausbreitung des Lichtes,'' (On the Influence of Gravitation on the Propagation of Light), \emph{Ann.\,d.\,Phys.} \textbf{35} 898--908 (1911). English translation on line at \url{https://einsteinpapers.press.princeton.edu/} , v.\,3, document 23, or in \emph{The Principle of Relativity: a collection of original papers}, trans.\ W.\,Perrett and G.\,B.\,Jeffrey (Dover Publications, NY, 1952) pp.\,99-108.\label{fn:Einstein1911}} 

\begin{figure}[h!]
\centering
\scalebox{0.9}{\includegraphics{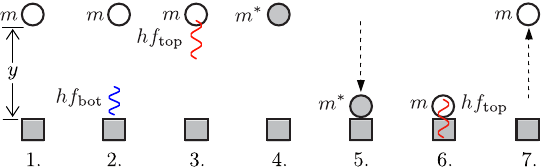}}
\caption{Derivation of the gravitational red shift}
\label{fig:red_shift_deriv}
\end{figure}

Consider a system of a mass $m$ and a reservoir of energy taken through the following series of steps (see Figure \ref{fig:red_shift_deriv}):
{\firmlists
\begin{enumerate}
\item The mass $m$ is at a height $y$ above a reservoir of energy.
\item The reservoir emits a photon of energy $hf_{\text{bot}}$ straight up, toward the mass.
\item The photon reaches the mass with an energy $hf_{\text{top}}$. 
\item This photon is absorbed by the mass. Its rest energy $mc^{2}$ is increased to $m^{*}c^{2} = mc^{2} + hf_{\text{top}}$.
\item The mass is lowered onto the reservoir as its potential energy $m^{\text{\textasteriskcentered}}gy$ is given to the reservoir.
\item The mass radiates an energy $hf_{\text{top}}$ into the reservoir and returns to its original value of $m$
\item A quantity of energy $mgy$ is extracted from the reservoir and used to lift the mass $m$ to its original height.
\end{enumerate}
}
The net change in the system's energy must be zero. As the net change in the mass's energy is zero, the net change in the reservoir's energy must be zero:
\begin{equation}
- hf_{\text{bot}} + m^{\text{\textasteriskcentered}}gy + hf_{\text{top}} - mgy = 0
\end{equation}
but
\begin{equation}
 m^{\text{\textasteriskcentered}} - m = \dfrac{hf_{\text{top}}}{c^{2}}
\end{equation}
so that, substituting this into the previous equation, 
\begin{equation}
- hf_{\text{bot}} + hf_{\text{top}}\dfrac{gy}{c^{2}} + hf_{\text{top}} = 0
\end{equation}
Dividing out by Planck's constant, and rearranging, leads to
\begin{equation}
f_{\text{bot}} = f_{\text{top}}\left(1 + \dfrac{gy}{c^{2}}\right)
\end{equation}
The frequency nearer the ground (where the gravitational potential is less) is higher than it is some distance above. Light moving up through a gravitational field loses energy, its frequency decreasing, and so is shifted toward the red end of the spectrum. This is sometimes (misleadingly) called the ``gravitational Doppler shift''. It isn't really a Doppler shift; that effect depends on the speeds of the wave, the source, and the receiver. In this example neither the source nor the receiver are moving. A better description of this phenomenon is the ``gravitational red shift'' or the ``gravitational frequency shift''. Had the light been emitted by the mass at $y$ and absorbed by the reservoir, the light would have been blue-shifted as it ``fell'' toward the reservoir. Note also that if $y$ is zero, then $f_{\text{top}} = f_{\text{bot}}$; there is no shift of frequency, no matter how large the value of $g$. This behavior is consistent with the clock hypothesis: \emph{local} acceleration has \emph{no} effect on an ideal clock's rate. 

The shift was confirmed by a brilliant experiment\footnote{R.\,V.\,Pound and G.\,A.\,Rebka, Jr., ``Gravitational red-shift in nuclear resonance'', \emph{Phys.\,Rev.\,Lett.} \textbf{3} (1959) 439--441.} about sixty-five years ago. A sample of a radioactive isotope of iron released gamma rays to be absorbed by a second sample of the same isotope some 22.5 meters vertically above the first. Because of the M\"{o}ssbauer effect, the frequency of gamma rays has to be the same (very nearly) to be re-absorbed by the second sample. As the gammas travelled upwards, their frequency decreased by a small amount, and were not absorbed. By moving the receiving iron \emph{downward} at a tiny speed (to undo the gravitational red shift with a compensating \emph{blue} Doppler shift) it was possible to adjust the gamma ray frequency (and at the same time, determine the shift to high precision) so that the radiation was now absorbed. The measured shift --- an astoundingly small effect, $gy/c^{2} = 2.45 \times10^{-15}$ --- agreed with the predicted value to within 10\%; further refinement\footnote{R.\,V.\,Pound and J.\,L.\,Snider, ``Effect of gravity on gamma radiation,'' \emph{Phys.\,Rev.}\,\textbf{140} (1965) B788-803.} found agreement to within 1\%.

Using the light's frequency as a metronome, the very same light in a given interval of time vibrates faster in a lesser gravitational potential than in a greater one. If a clock is calibrated with the frequency of this light, the period of the vibrations are related by the inverse relation as was found above, because $f = 1/T$, where $T$ is the period. That is,
\begin{equation}
T_{\text{top}} = T_{\text{bottom}}\left(1 + \dfrac{gy}{c^{2}}\right)
\end{equation}
The period is the interval of time between vibrations. There is nothing special about that particular interval. It must be that for \emph{any} time interval $\Delta t$ between two events, the time interval $\Delta t_{\text{top}}$ higher in the gravitational potential and the corresponding interval lower in the potential are related in the same way:
\begin{equation}
\Delta t_{\text{top}} = \Delta t_{\text{bottom}}\left(1 + \dfrac{gy}{c^{2}}\right) %
\end{equation}
In short, \emph{clocks run more slowly} where the gravitational potential is less: gravitational time dilation.\footnote{Imagine two jars filled with cookies, one jar containing more cookies. If the cookies are all the same size, one jar must be larger; if the jars are the same size, the cookies in one must be smaller. Consider a number of light waves emitted with frequency $f_{2}$ at a potential $S_{2}$ and propagating to a lower one, $S_{1}$, arriving with frequency $f_{1} > f_{2}$. The same number of vibrations must arrive at $S_{1}$ as were emitted at $S_{2}$ per unit \emph{local} time. (``If there is constant transmission of light from  $S_{2}$ to $S_{1}$, how can any other number of periods per second arrive in $S_{1}$ than is emitted in $S_{2}$?'' Einstein, note \ref{fn:Einstein1911}, p.\,105.) Let this number be analogous to the size of the cookies, and the seconds to jars. As the vibrations per local second are the same, the seconds are effectively larger in the lower potential. (Alas, another twin puzzle: this analogy is from Alex Boyer or her twin sister Vicki, class discussion, 1996; author memory uncertain. ``The students are in schools to teach the teachers,'' J.\,A.\,Wheeler.)} Distant observers would perceive that the flow of time slows to a stop at the event horizon of a black hole; Russian relativists formerly called black holes \emph{frozen stars}. This effect will provide a second resolution of the twins.

Again assume that the time of acceleration is small (though not necessarily zero). Bill's accounting of the times is the same as before. During her trip to Proxima, Alice believes that Bill's clock moves from zero to 1.8 years, while hers moves from zero to 3 years. Denote by $\Delta t_{\text{acc}}$ the time of the acceleration during which Bill reverses velocity as measured by Alice. She cannot distinguish the effects of her acceleration $a$ from the effects of gravity, and so one presumes that the effect of her acceleration is just the same as if she had been in a uniform gravitational field, which she imagines extends from her position to Bill's. Had Alice's acceleration taken place near Bill, it would have had very little effect: the gravitational potential of a uniform field increases linearly with distance; zero distance, zero difference in clock rate. From her point of view, during her acceleration Bill is at a higher gravitational potential, and consequently she ascribes to Bill's time $\Delta t_{\text{acc}}'$ the value
\begin{equation}
\Delta t_{\text{acc}}' = \Delta t_{\text{acc}}\Bigl(1 + \dfrac{aL}{c^{2}}\Bigr) = \Delta t_{\text{acc}} + \dfrac{a \Delta t_{\text{acc}}L}{c^{2}}
\end{equation}
Whatever its size, $\Delta t_{\text{acc}}$ has to satisfy the condition (to lowest order in $(v/c)^{2}$)
\begin{equation}
a \Delta t_{\text{acc}} = 2v
\end{equation}
to reverse Alice's velocity from $\mathbf{v}$ to $-\mathbf{ v}$. If $a$ is sufficiently large, $\Delta t_{\text{acc}}$ is negligible in comparison with $a\Delta t_{\text{acc}}$, and may be ignored. During the turnaround, because of the equivalence principle and the gravitational red shift, Alice observes an increase in Bill's clock time of 
\begin{equation}\label{eq:Born_term}
\dfrac{a \Delta t_{\text{acc}}L}{c^{2}} = \dfrac{2vL}{c^{2}}
\end{equation}
which is just the same amount as with the instantaneous change in coordinate systems. The total time Alice ascribes to Bill's clock when she returns is
\begin{equation}
1.8\,\text{yr} + \dfrac{2(0.8\,c)(4\,\text{ly})}{c^{2}} + 1.8\,\text{yr} = 10\,\text{yr}
\end{equation}
as required. Once again, the ``paradox'' is resolved. The calculation was first presented in a popular treatment of relativity by Einstein's good friend Max Born in 1920.\footnote{Max Born, \emph{Einstein's Theory of Relativity}, (Dover Publications, New York, 1962, 1965), pp.\,354--356; this is a revised and updated English version of a book published in Berlin in 1920. Perhaps following Einstein (note (\ref{fn:Einstein_dialog}), Born describes this approximate resolution as based on general relativity. Like Tolman (note \ref{fn:Tolman}), Born claims that the exact resolution requires a full general relativistic treatment. An essentially identical solution appears in V.\,Fock's \emph{The Theory of Space, Time, and Gravitation} (Pergamon Press, New York, 1955, trans.\ N.\,Kemmer), pp.\,211--214. Note Fock's last sentence on p.\,214: ``We have performed the calculation in the inertial frame of reference. To repeat it in a frame connected with [the accelerated clock] would have no sense for we would only be evaluating the same integrals in terms of different variables.'' Few doubt that the twins must agree, but it is worth looking at \emph{how} they agree. As this paper shows, Fock's claim is correct, but only if applied to the \emph{sum} of integrals for the complete journey; the twins do \emph{not} agree for the times arising from individual parts of the trajectory.\label{fn:Born}}  A similar solution was provided by R.\ C.\ Tolman\footnote{R.\,C.\,Tolman, \emph{Relativity, Thermodynamics and Cosmology}, (Dover Publications, New York, 1987; reprint of 1934), pp.\,194--197. Tolman states at the end of his solution, ``The treatment of the problem without approximation would involve the full apparatus of the general theory of relativity.''\label{fn:Tolman}} in 1934.

Some might say that invoking the equivalence principle, exemplified by the gravitational red shift, to reconcile the twins' times is evidence that only general relativity can resolve the difficulty. Indeed, the gravitational red shift was one of the three classic tests of general relativity proposed by Einstein\footnote{A.\,Einstein, ``Die Grundlage der allgemeinen Relativit\"{a}tstheorie'' (The Foundation of the General Theory of Relativity), \emph{Ann.\,d.\,Phys.} \textbf{49} 769--822, (1916). English translation in Perrett and Jeffrey (see \ref{fn:Einstein1911}), pp.\,111--164; the red shift is discussed on pp.\,161--162. This epochal paper is a compendium of a series of lectures Einstein gave in November, 1915 to the Prussian Academy of Sciences, Berlin. Strangely, the first page of the article is not published in Perrett and Jeffrey; it may be found in the online collection.} in 1915. In general relativity, gravitational time dilation comes directly from the square root of the component $g_{00}$ of the metric tensor. The component is, to a first approximation, given by 
\begin{equation}
g_{00} = 1 + \dfrac{2\varphi}{c^{2}}
\end{equation}
where $\varphi = -GM/r$ is the gravitational potential for a spherical mass distribution. Near the Earth, $\varphi = U_{\text{grav}}/m = gy$. (The standard Newton expression for gravitational force comes out of Einstein's equations by differentiation\footnote{L.\,D.\,Landau and E.\,M.\,Lifshitz, \emph{The Classical Theory of Fields}, 3rd rev.\,ed., (Pergamon Press, London and Addison-Wesley, Reading, MA, 1971), \S96, ``Newton's Law'', pp.\,278--279.} of $g_{00}$.) It was later realized however that, as shown above, the gravitational red shift does \emph{not} require general relativity; it is as a consequence of the equivalence principle and special relativity. Einstein had in fact derived the effect in 1911, four years before he arrived at general relativity (note \ref{fn:Einstein1911}). Imagine a photon moving from ceiling to floor in an accelerating elevator. By working out the Doppler shift, one obtains a frequency shift proportional to the acceleration and the height of the elevator. In his magisterial biography of Einstein, Pais\footnote{A.\,Pais, \emph{Subtle is the Lord...\,The Science and Life of Albert Einstein}, (Oxford U.\ Press, 1982), pp.\,197--198. Pais' biography, unlike most, is rich in the actual physics. As the mathematicians might say, the red shift is \emph{necessary} but not \emph{sufficient} to affirm general relativity.} remarks that
\begin{quote}
[I]n good texts on general relativity the red shift is taught twice. In a first go-around, it is noted that the red shift follows from special relativity and the equivalence principle only. Then, after the tensor equations of general relativity have been derived and the equivalence principle has been understood to hold strictly only in the small, the red shift is returned to and a proof is given that it is sufficient for the derivation of the previous result to consider only the leading deviations of $g_{00}$ [Pais writes $g_{44}$] from its flat-space-time value.\footnote{It might occur to the reader that Pais's remark about the equivalence principle holding strictly in the small would argue against applying it to twins separated by light-years. Pais correctly limits the principle to small regions of spacetime where there is a genuine gravitational field, with nonzero curvature; only locally can this curvature be regarded as flat Minkowski space. There is \emph{no} curvature in the accelerating twin's journey (apart from that due to the Earth and Sirius, which is ignored) and thus no restriction to small sections of spacetime. It is the principle's basis on flat spacetime that prevents its being used to derive the correct bending of light near a star, the Einstein value of $4GM/Rc^{2}$, rather than the Newtonian value, half as large, which results from the equivalence principle. The curvature of space produces the other half.}
\end{quote}

\section{Special relativistic variation on the M\o ller-Arz\`{e}lies solution}

A complete, detailed solution making use of general relativity, particularly the metric tensor, was given by M\o ller long ago.\footnote{\label{fn:Moller_Lorentz}C.\,M\o ller, ``On homogeneous gravitational fields in the general theory of relativity and the clock paradox'', \emph{Mat.-Fys.\ Medd.\ Dan.\ Vid.\ Selks.}\ \textbf{20} (19) (1943) 1--24. M\o ller's coordinates were based on a similar set introduced by Lorentz: H.\,A.\,Lorentz, \emph{Das Relativitätsprinzip}, B.\,G.\,Teubner, Berlin, 1914.} He revisited his solution, still within the framework of general relativity, in his famous treatise,\footnote{C.\,M\o ller, \emph{The Theory of Relativity} 2nd ed., Oxford U. Press, (1972), pp.\,292-296.\label{fn:Moller_Rel}} but his conclusion was excessively modest. Having done the hard work, he observed that the ``paradox'' was resolved in the limit as the traveler's proper (constant) acceleration approached an infinite value. M\o ller's result was subsequently discussed by Arz\`{e}lies\footnote{H.\,Arz\`{e}lies, \emph{Relativité Géneralisée Gravitation}, Gauthier-Villars, Paris, (1961), v.1, pp.\,317--323. In \emph{Relativistic Kinematics} (note \ref{fn:ArzeliesSpecial}), he writes ``The general theory, which is beyond the scope of the present work, provides the only means of analysing all the features of the phenomenon,'' (p.\,188). Concerning Tolman's solution he states ``I cannot agree with this author that general relativity is necessary to obtain a solution of the problem,'' (p.\,192).\label{fn:ArzeliesGeneral}} who showed explicitly the equality between values obtained by the two twins for \emph{any} value (except zero) of the (constant) acceleration. It is perhaps worth remarking that Arzeliès chose to discuss the twins not in either of the two books he wrote on relativistic mechanics,\footnote{\label{fn:ArzeliesSpecial}H.\,Arz\`{e}lies, \emph{La Cinématique relativiste},  Gauthier-Villars, Paris, (1955), English translation \emph{Relativistic Kinematics}, Pergamon Press, New York, (1965); H.\,Arz\`{e}lies, \emph{La Dynamique relativiste et ses applications}, Gauthier-Villars, Paris, v.1, (1957), v.2, (1958), or in English, \emph{Relativistic Point Dynamics}, trans.\,P.\ W.\ Hawkes, Pergamon Press, New York, (1972). Alas, these are the only two of Arz\`{e}lies's seven books on relativity that have been translated into English.} but in his text on general relativity. The remainder of this paper is devoted to a special relativistic version of the M\o ller-Arz\`{e}lies solution. The calculations for this third solution are lengthier than those for the previous two resolutions, but the steps are simple. To begin, relativistic motion with constant acceleration will be reviewed to obtain some necessary formulas. Next, the \emph{M\o ller coordinates} will be introduced. These allow one to transform from inertial to constantly accelerated reference frames. The relation between $X$, the M\o ller position and the corresponding time $T$ in these coordinates will be derived from a well-known result in the calculus of variations. Finally the travel time for each twin as measured by Bill and then by Alice will be calculated. The calculus turns out to be surprisingly straightforward.

\subsection{Proper time recorded by an accelerating clock}

So long as two reference frames $S$ and $S'$ are in uniform relative motion with respect to each other, the usual Lorentz transformations may be used to determine measurements in one reference frame in terms of the other. With a frame $S''$ accelerating with respect to frame $S$, things are not so simple. One approach is to use, in addition to $S$, a \emph{family} of inertial frames $S_{i}'$, one of which at any given moment is momentarily at rest with respect to $S''$, and in uniform relative motion to $S$. There is no difficulty moving between any one of the $S'_{i}$ and $S$ with Lorentz transformations. The differential $d\tau$ of an object's proper time $\tau$ in $S$ is given as usual by
\begin{equation}
\begin{aligned}
d\tau &= \sqrt{dt^{2} - (dx^{2} + dy^{2} + dz^{2})/c^{2}}\\
&= \sqrt{1 - (v/c)^{2}}\,dt = dt/\gamma
\end{aligned}
\end{equation}
with $\gamma$ taking its usual form. Define in the usual way the four velocity $u^{\mu}$ and the four acceleration $a^{\mu}$;
\begin{equation}
u^{\mu} = \dfrac{dx^{\mu}}{d\tau}\,; \qquad a^{\mu} = \dfrac{du^{\mu}}{d\tau}
\end{equation}
All the motions to be considered here will be taken along an axis, specifically the $x$-axis, in which case
\begin{equation}\label{eq:4-velocity}
(u^{0}, u^{1}, u^{2}, u^{3}) = (\gamma c, \gamma v, 0, 0)
\end{equation}
and 
\begin{equation}\label{eq:a_mu}
\begin{aligned}
(a^{0}, a^{1}, a^{2}, a^{3}) &= \gamma (c d\gamma/dt, d(\gamma v)/dt, 0, 0)\\
 &= \gamma^{4}(v/c, 1, 0, 0)\,dv/dt
\end{aligned}
\end{equation}
From (\ref{eq:a_mu}) it follows that $a^{0} = (v/c)a^{1}$, and
\begin{equation}\label{eq:square_of_four-accel}
a^{\mu}a_{\mu} = - (a^{1}/\gamma)^{2} = - (d(\gamma v)/dt)^{2} 
\end{equation}
Let an object with coordinates $(ct, x, 0, 0)$ in the inertial frame $S$ be moving along the $x$-$x'$ axis with speed $v = dx/dt$ with respect to the $S$ frame. Let the inertial frame $S'$ be instantaneously at rest with respect to this object. In the $S'$ coordinates, the object's location is given by $(ct', x', 0, 0)$ and its speed $v' = dx'/dt'$ is (instantaneously) equal to zero. Suppose that the object experiences an acceleration relative to $S'$ along the $x$-$x'$ axis equal to a constant value, $g$, as measured in the $S'$ frame. In that frame the nonzero components of the acceleration are
\begin{equation}
(a^{\prime\,0}, a^{\prime\,1}) = (0, dv'/dt') = (0, g)
\end{equation}
By Lorentz invariance, 
\begin{equation}
a^{\mu}a_{\mu} = a^{\prime\,\mu}a^{\prime}_{\mu} = - g^{2}
\end{equation}
and so from (\ref{eq:square_of_four-accel})
\begin{equation}
\dfrac{d(\gamma v)}{dt} = g
\end{equation}
If the initial velocity of the object is zero with respect to $S$, 
\begin{equation}\label{eq:t_accel}
\gamma v = gt
\end{equation}
Solving for $v$,
\begin{equation}\label{eq:v_acc(t)}
v = \dfrac{gt}{\sqrt{1 + (gt/c)^{2}}}
\end{equation}
and 
\begin{equation}\label{eq:gamma_acc(t)}
\gamma = \sqrt{1 + (gt/c)^{2}}
\end{equation}
The distance $x$ traveled in the $S$ frame by the object in time $t$ is 
\begin{equation}\label{eq:position_time_Bill}
x = \int v\,dt = (c^{2}/g)\sqrt{1 + (gt/c)^{2}}\, + \mbox{ constant}
\end{equation}
If the initial position of the object is the origin, the constant becomes $-(c^{2}/g)$, and
\begin{equation}\label{eq:position_accel}
x = (c^{2}/g)\bigl[\sqrt{1 + (gt/c)^{2}} - 1\bigr] = (c^{2}/g)[\gamma - 1]
\end{equation}
Note that for $v/c \ll 1$, $gt/c \ll1$, $\gamma \approx 1 + \tfrac{1}{2}(v/c)^{2}$, and the two quantities in (\ref{eq:position_accel}) reduce to the usual Galilean relations $x = \tfrac{1}{2}gt^{2}$ and $x = v^{2}/(2g)$, respectively. The worldline of the object is thus described by the hyperbola
\begin{equation}
\bigl[x + (c^{2}/g)\bigr]^{2}  - (ct)^{2} = \bigl(c^{2}/g\bigr)^{2}
\end{equation}
Motion with constant proper acceleration is often called ``hyperbolic''.\footnote{Arz\`{e}lies (note \ref{fn:ArzeliesGeneral}), p.\,324 states that the term ``hyperbolic motion'' is due to M.\ Born, citing his article ``Die Theorie des starren Elektrons in der Kinematic des Relativitäts-Prinzipes'' (The theory of the rigid electron in the kinematics of the principle of relativity), \emph{Ann.\,d.\,Phys.} \textbf{30} (11) 1--56 (1909).} An observer in hyperbolic motion whose clock was synchronized with the $S$ clock at the beginning of her motion will record a time $\tau$ on her clock given by
\begin{equation}\label{eq:proper_time_accel}
\tau = \int \dfrac{1}{\gamma}\,dt = (c/g)\,\text{arc sinh}(gt/c)
\end{equation}
This is her proper time. Inverting this relation gives
\begin{equation}\label{eq:t_in_terms_of_tau_accel}
t = (c/g) \sinh (g\tau/c)
\end{equation}
Plugging this into (\ref{eq:v_acc(t)}),
\begin{equation}\label{eq:v_acc(tau)}
v(\tau) = c \tanh (g\tau/c)
\end{equation}
From (\ref{eq:gamma_acc(t)}) or from (\ref{eq:v_acc(tau)})
\begin{equation}\label{eq:gamma_acc(tau)}
\gamma(\tau) = \sqrt{1 + \sinh^{2}(g\tau/c)} = \cosh(g\tau/c)
\end{equation}
so from (\ref{eq:position_accel})
\begin{equation}\label{eq:x_in_terms_of_tau_accel}
x(\tau) = (c^{2}/g)\bigl[\cosh(g\tau/c) - 1\bigr]
\end{equation}
These formulas are all that one needs to work out Alice's time as she measures it, and as Bill measures it.

\subsection{The M\o ller coordinates}

To calculate Alice's reckoning of Bill's age, it's necessary to transform from Alice's rest frame, $S''$ to Bill's, $S$ during the periods of acceleration. The appropriate formulas are often known as M\o ller coordinates, as he derived them in his solution to the twins (note \ref{fn:Moller_Rel}, pp.\,287--290). The simplest approach to the M\o ller coordinates may be as described by Misner, Thorne and Wheeler.\footnote{C.\,W.\,Misner, K.\,S.\,Thorne, and J.\,A.\,Wheeler, \emph{Gravitation}, (W.\,H.\,Freeman \& Co., San Francisco, 1971); reissued by Princeton U.\,P., Princeton, (2017), pp.\,169--174. Another very full discussion is given by F.\,W.\,Sears and R.\,W.\,Brehme, \emph{Introduction to the Theory of Relativity}, (Addison-Wesley, Reading, MA, 1968), pp.\,70--74. See also R.\,A.\,Mould, \emph{Basic Relativity}, (Springer-Verlag, New York, 1994), Chapter 8, ``Uniform Acceleration'', pp.\,221--268, and Appendix E, ``Uniformly Accelerated Transformation Equations'', pp.\,432--434.\label{fn:MTW}}

Consider an event $\mathcal{P}(\tau)$ on the worldline of an object with constant nonzero proper acceleration $g$ in the direction of the $x$ axis in the $S$ frame. Say that the object set out from the origin with zero initial velocity.  Let $\mathcal{P}(\tau)$'s coordinates in the $S$ frame be denoted $z^{\mu}(\tau)$. Explicitly, from (\ref{eq:t_in_terms_of_tau_accel}) and (\ref{eq:x_in_terms_of_tau_accel}) 
\begin{equation}\label{eq:z(tau)}
\begin{aligned}
z^{0} &= (c/g) \sinh (g\tau/c)\\
z^{1} &= (c^{2}/g)\bigl[\cosh(g\tau/c) - 1\bigr]\\
z^{2} &= z^{3} = 0
\end{aligned}
\end{equation}
One point does not a coordinate system make. To provide for the other points in the $S''$ frame, construct at $\mathcal{P}(\tau)$ a set of three orthonormal (in the Minkowski sense) four-vectors, $\mathsf{e}^{\mu}_{i}(\tau)$, $i = 1,2,3$. A fourth orthonormal vector, $\mathsf{e}^{\mu}_{0}(\tau)$, is tangent to the object's worldline, parallel to the object's 4-velocity $u^{\mu}$. Using (\ref{eq:4-velocity}), (\ref{eq:gamma_acc(tau)}), and (\ref{eq:v_acc(tau)}), 
\begin{equation}\label{eq:u(tau)}
u^{\mu} =  (\gamma c, \gamma v, 0, 0) = (c \cosh (g\tau/c), c \sinh (g\tau/c), 0, 0)
\end{equation}
which leads to the identification
\begin{equation}
\begin{aligned}\label{eq:hyperplane_basis}
\mathsf{e}^{\mu}_{0}(\tau) &= (\cosh (g\tau/c), \sinh (g\tau/c), 0, 0)\\
\mathsf{e}^{\mu}_{1}(\tau) &= (\sinh (g\tau/c), \cosh (g\tau/c), 0, 0)\\
\mathsf{e}^{\mu}_{2}(\tau) &= (0, 0, 1, 0)\\
\mathsf{e}^{\mu}_{3}(\tau) &=(0, 0, 0, 1)
\end{aligned}
\end{equation}

The point $\mathcal{P}$ on the worldline, together with the three 4-vectors $\{\mathsf{e}^{\mu}_{1}, \mathsf{e}^{\mu}_{2}, \mathsf{e}^{\mu}_{3}\}$ define a hyperplane orthogonal to the worldline. That is, the location of any point  $\mathcal{Q}$ on the worldline can be described by $x^{\mu}(\tau)$ in the frame $S$ by the expression (see Fig.\,4)
\begin{equation}\label{eq:vect_add_x_z}
x^{\mu}(\tau) = z^{\mu}(\tau) + X\mathsf{e}^{\mu}_{1}(\tau) + Y\mathsf{e}^{\mu}_{2}(\tau) + Z\mathsf{e}^{\mu}_{3}(\tau)
\end{equation}
The numbers $X, Y, Z$ serve as Minkowski coordinates for any given point $\mathcal{Q}$ on the worldline. As time moves forward, so does the hyperplane, at some time $\tau$ cutting through the point $\mathcal{Q}$. Identify the fourth coordinate $T$ with this value of $\tau$, i.e.,
\begin{equation}
X^{\mu} = (cT, X, Y, Z)
\end{equation}
From (\ref{eq:z(tau)}), (\ref{eq:hyperplane_basis}) and (\ref{eq:vect_add_x_z}) come the M\o ller coordinates, 
\begin{equation}\label{eq:Moller_coords}
\begin{aligned} 
t &= \left[\dfrac{c}{g} + \dfrac{X}{c}\right] \sinh (gT/c)\rule[-3ex]{0pt}{3ex}\\
x &= \left[\smash[t]{\dfrac{c^{2}}{g} }+ X\right] \cosh(gT/c) - (c^{2}/g)\rule[-3ex]{0pt}{3ex}\\
y &= Y\rule[-1.5ex]{0pt}{1.5ex}\\
z &= Z
\end{aligned}
\end{equation}
\begin{figure}[h!]
\centering
\scalebox{0.9}{\includegraphics{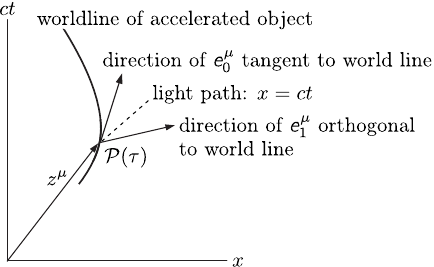}}
\caption{The geometry of the accelerated frame\\ (adapted from MTW, note \ref{fn:MTW}, Fig.\,6.3, p.\,172.)}
\label{fig:MTW}
\end{figure}
\indent The expression for the differential of the proper time is given in terms of the $X^{\mu}$ coordinates by\footnote{Both M\o ller and Arz\`{e}lies derive (\ref{eq:proper_time_int}) from general relativistic considerations. The M\o ller coordinates can also be derived by requiring that the Minkowski proper time $d\tau$ be transformed into (\ref{eq:proper_time_int}) following the method in Lorentz (note \ref{fn:Moller_Lorentz}).}
\begin{gather}
d\tau =\, \sqrt{dt^{2} - \frac{1}{c^{2}}(dx^{2} + dy^{2} + dz^{2})} \notag \\
\label{eq:proper_time_int} \;=\, \sqrt{\left(1 + \frac{gX}{c^{2}}\right)^{\!2} \!- \left(\dfrac{V}{c}\right)^{\!2}}\,dT
\end{gather}
where, as motion is restricted to the $x$-$X$ axis,
\begin{equation}
V = dX/dt\,; \quad dY/dt = dZ/dt = 0
\end{equation}
\indent One may invert the M\o ller coordinates algebraically to obtain expressions for $X$ and $T$ in terms of $x$ and $t$. That procedure does not, unfortunately, give the necessary relations between $X$ and $T$. A variational principle and the equivalence principle will provide them.

\subsection{Determining $X(T)$ and $V(T)$}

From the point of view of Alice, the acceleration she ascribes to Bill's frame of reference will be interpreted as \emph{free fall} within a constant gravitational field. While her earthbound brother is in free fall, Alice remains at her own origin, because in her rest frame $S''$ the rocket engines exactly counteract the gravitational force. The free fall naturally maximizes the integral\footnote{An observer's free fall maximizes her proper time. E.\,F.\,Taylor calls this result ``the Principle of Maximal Aging'': E.\,F.\,Taylor, J.\,A.\,Wheeler, and E.\,F.\,Bertschinger, \emph{Exploring Black Holes: Introduction to General Relativity}, 2nd ed., p.\,1-13. Available for free from Taylor's site, \url{https://www.eftaylor.com} , or from the Internet Archive, \url{https://ia803109.us.archive.org/23/items/exploringblackholes/EBH-2.pdf} .}
\begin{equation}
\int d\tau = \int \sqrt{\left(1 + \dfrac{gX}{c^{2}}\right)^{\!2} - \left(\dfrac{V}{c}\right)^{\!2}}\,dT \equiv \int \mathcal{L}\;dT
\end{equation}
Since the integrand $\mathcal{L}$ does not depend explicitly on the variable $T$, there is a constant ``first integral'', $K$, (analogous to energy)
\begin{equation}
K = V\dfrac{\partial \mathcal{L}}{\partial V} - \mathcal{L}  = - \dfrac{1}{\mathcal{L}}\,\left(1 + \dfrac{gX}{c^{2}}\right)^{\!2}
\end{equation}
This can be solved for $V^{2}$ and then $V$:
\begin{equation}\label{eq:V(T)}
V = \pm\,(c/K)\Bigl(1 + \dfrac{gX}{c^{2}}\Bigr)\sqrt{K^{2} - \Bigl(1 + \dfrac{gX}{c^{2}}\Bigr)^{\!2}}
\end{equation}
(One should allow for either sign.) Because $V = dX/dT$, (\ref{eq:V(T)}) is a differential equation for $X(T)$: 
\begin{equation}
\dfrac{dX}{dT} = \pm\,c\Bigl(1 + \dfrac{gX}{c^{2}}\Bigr)\sqrt{1 - \dfrac{1}{K^{2}}\Bigl(1 + \dfrac{gX}{c^{2}}\Bigr)^{\!2}}
\end{equation}
Setting
\begin{equation}\label{eq:u(T)}
u = \dfrac{1}{K}\Bigl(1 + \dfrac{gX}{c^{2}}\Bigr)
\end{equation}
the differential equation becomes
\begin{equation}
(c/g) \dfrac{du}{dT} = \pm\,u\sqrt{1 - u^{2}}
\end{equation}
The equation is separable into a standard form:\footnote{Note $du/(u\sqrt{1 - u^{2}}) = - d\,\mathrm{arc\,sech}(u)$.}
\begin{equation}
\begin{aligned}
(g/c)\int dT &= \pm\,\int \dfrac{du}{u\sqrt{1 - u^{2}}}\\
(g/c)(T - T_{0}) &= \mp\,\mbox{arc sech}\left[\dfrac{1}{K}\Bigl(1 + \dfrac{gX}{c^{2}}\Bigr)\right]
\end{aligned}
\end{equation} 
(the quantity $(g/c)T_{0}$ is a constant of integration). The function $\mbox{arc sech}(x)$ is even, so the sign ambiguity is lost. Solving for $X(T)$, 
\begin{equation}\label{eq:X(T)}
X(T) = (c^{2}/g) \left[K\,\mbox{sech} \left((g/c)(T - T_{0})\right) - 1\right]
\end{equation}
$K$ will be determined from the initial conditions: the values of $T_{0}$ and $X(T_{0})$. That's all one needs to work out Bill's time from Alice's point of view.

\subsection{The travel times according to Bill}

\begin{figure}[h!]
\centering
\scalebox{1.33}{\includegraphics{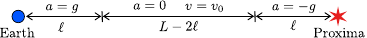}}
\caption{The flight plan (in the Earth-Proxima frame)}
\label{fig:flight_plan}
\end{figure}

Alice needs a flight plan for her round trip to the star's neighborhood. Taking all measurements within the Earth-Proxima frame, let the distance from Earth to the turnaround point be denoted $L$. Let Alice leave Earth traveling with a constant proper acceleration $g$ until she attains a speed $v_{0}$, covering a distance $\ell$ during that time. Attaining that speed, she shuts off her engines and coasts at a speed $v_{0}$ until she is the same distance, $\ell$, from her closest approach to Proxima. She then turns on her retrorocket engines, experiencing a proper acceleration $-g$, momentarily achieving a speed of zero at her closest approach to the star. The acceleration reverses her velocity back towards Earth, reaching that speed again at a distance $\ell$ from her closest approach. Again she shuts down her engines, coasts until she is $\ell$ from Earth, and finally turns on her retrorockets to bring her velocity to zero at the Earth's surface. The journey as Bill reckons it is represented schematically in Fig.\,\ref{fig:flight_plan}.

To minimize confusion, say that the time Bill measures for Alice's trip is $\Delta t_{\text{Bill}}$ on his own clock, and $\Delta t_{\text{Alice}}'$ on her clock; similarly let the time Alice measures be $\Delta t_{\text{Alice}}$ on her own clock, and $\Delta t_{\text{Bill}}'$ on his clock; the primes indicating that these times are what one twin measures on the \emph{other} twin's clock. Extend this convention to all measurements: no prime on one's own proper measurements, primes on the other as observed by the first. The resolution of the twins lies in demonstrating that 
\begin{equation}
\Delta t_{\text{Bill}} = \Delta t_{\text{Bill}}'\,; \quad \Delta t_{\text{Alice}} = \Delta t_{\text{Alice}}'
\end{equation}
In words, when they are reunited at the end of Alice's journey, the time Bill measures on his clock must be the time Alice measures on Bill's clock; the time Alice measures on her clock must be the time Bill measures on Alice's clock. The calculations will be shown in some detail, because, as Arz\`{e}lies says (see note \ref{fn:ArzeliesGeneral}, p.\,322), \emph{il y a beacoup de sceptiques}: there are many who are skeptical that there is no problem.

As measured by Bill, it takes Alice a distance (\ref{eq:position_accel})
\begin{equation}\label{eq:ell}
\ell = (c^{2}/g) \left[\gamma - 1\right]
\end{equation}
to accelerate from $v = 0$ to $v = v_{0}$ in a time (\ref{eq:t_accel})
\begin{equation}\label{eq:t_0}
t_{0} = \dfrac{\gamma v_{0}}{g} \equiv t_{\text{B}}
\end{equation}
By symmetry, Bill will measure the same distance and the same times for all of Alice's accelerations from to $v = 0$ to $v_{0}$ or back, so the total time $t_{\text{accel}}$ he measures for Alice's acceleration is $4t_{\text{B}}$. The remainder of the travel time Alice spends coasting a distance $2(L - 2\ell)$ at a speed $v_{0}$. The total time $\Delta t_{\text{Bill}}$ believes Alice's trip takes by his own clock is then given by $t_{\text{coast}} + 4t_{\text{B}}$, i.e., 
\begin{equation}\label{eq:t_Bill}
\Delta t_{\text{Bill}} = t_{\text{coast}} + t_{\text{accel}} =  \dfrac{2L}{v_{0}} - \dfrac{4\ell}{v_{0}} + \dfrac{4\gamma v_{0}}{g}
\end{equation}
It's a little more transparent to write the second term without reference to $\ell$, using (\ref{eq:ell}):
\begin{equation}\label{eq:t_Bill_no_ell}
\Delta t_{\text{Bill}} = \dfrac{2L}{v_{0}} + \dfrac{4c^{2}}{v_{0}g} \left[1 - \dfrac{1}{\gamma}\right]
\end{equation}
\indent What time does Bill measure on Alice's clock? For the coasting time, Alice's clock will be observed to run at a rate of $(1/\gamma)$ of Bill's, in accord with special relativity. The time $t_{0}'$ Bill ascribes to Alice's clock as she accelerates from $v = 0$ to $v = v_{0}$ is the same time Alice measures on her own clock (it is just her proper time, $\tau$), given by (\ref{eq:proper_time_accel})
\begin{equation}\label{eq:Alice_proper_time_accel}
\begin{aligned}
\tau &= t_{0}' = (c/g)\,\text{arc sinh}(gt_{0}/c)\\
&= (c/g)\,\text{arc sinh}(\gamma v_{0}/c)\\
&=  (c/g)\,\text{arc sinh}(t_{\text{B}}) \equiv \tau_{A}
\end{aligned}
\end{equation}
Again by symmetry, the total accelerating time on Alice's clock according to Bill is just $4\tau_{A}$. Then $\Delta t_{\text{Alice}}' = (1/\gamma)t_{\text{coast}} + 4\tau_{A}$, i.e.,
\begin{equation}\label{eq:tprime_Alice}
\Delta t_{\text{Alice}}' =  \dfrac{2L}{\gamma v_{0}} - \dfrac{4\ell}{\gamma v_{0}} + \dfrac{4c}{g}\,\text{arc sinh}(\gamma v_{0}/c)
\end{equation}
or equivalently,\footnote{Note $\text{arc sinh}(x) = \ln(x + \sqrt{1 + x^{2}})$, and $1 + \gamma^{2} (v_{0}/c)^{2} = \gamma^{2}$.}
\begin{equation}\label{eq:tprime_Alice_no_ell}
\Delta t_{\text{Alice}}' =  \dfrac{2L}{\gamma v_{0}} - \dfrac{4\ell}{\gamma v_{0}} + \dfrac{4c}{g}\ln\left(\gamma\left[1 + \dfrac{v_{0}}{c}\right]\right)
\end{equation}
\indent What are the numerical values of these times? It is a happy coincidence that if $g$ has a value equal to the Earth's gravitational field, say 9.8 m/s$^{2}$, the expression $c^{2}/g$ is very nearly equal to a light-year, and $c/g$ to a year. The value of $g$ required to make them equal is 9.5 m/s$^{2}$. Let Alice's spaceship accelerate at just this value, to a maximum speed of $v_{0} = 0.8 c$, and let $L = 4$ light-years as in the earlier versions of Alice's journey. Then $\gamma = \tfrac{5}{3}$, $\ell = (c^{2}/g)[\gamma - 1] = \tfrac{2}{3}$ ly, and 
\begin{equation}\label{eq:t_Bill_values}
\begin{aligned}
\Delta t_{\text{Alice}}' &= \dfrac{2\cdot 4\,\text{ly}}{\tfrac{5}{3}\!\cdot \tfrac{4}{5}c} -  \dfrac{4\cdot \tfrac{2}{3}\,\text{ly}}{\tfrac{5}{3}\cdot \tfrac{4}{5}c} + 4(c/g)\ln\left(\tfrac{5}{3}\!\left[1 + \tfrac{4}{5}\right]\right)\rule[-3.5ex]{0pt}{3.5ex}\\
&= (6 - 2 + 4\ln 3)\,\mbox{yr} = 8.39\,\mbox{years}\rule[-2ex]{0pt}{2ex}\\
\Delta t_{\text{Bill}} &= \dfrac{2\cdot 4\,\text{ly}}{\tfrac{4}{5}c} - \dfrac{4\cdot\tfrac{2}{3}\,\text{ly}}{\tfrac{4}{5}c} + 4(c/g)\bigl(\tfrac{5}{3}\cdot \tfrac{4}{5}\bigr)\rule[-3.5ex]{0pt}{3.5ex}\\\
&= (10 - \tfrac{10}{3} + \tfrac{16}{3})\,\mbox{yr} = 12\,\mbox{years}\\
\end{aligned}
\end{equation}
For comparison, the Born \emph{et al.}\ equivalence principle solutions and the Helliwell \emph{et al.}\ change in frame solutions give the same answers as the naive view, but with the twins' clocks reconciled: 
\begin{equation}\label{eq:approx_times}
\begin{aligned}
\Delta t_{\text{Alice, approx}} &= \Delta t'_{\text{Alice, approx}} = 6\,\text{years};\\
\Delta t_{\text{Bill, approx}} &= \Delta t'_{\text{Bill, approx}}= 10\,\text{years}
\end{aligned}
\end{equation}

\subsection{The travel times according to Alice}

From Alice's point of view, the Earth-Proxima frame $S$ moves. During periods of no acceleration the frame $S$ goes by Alice at a constant speed $v_{0}$. The distance which Bill measures to be $L - 2\ell$ will be measured as $(L - 2\ell)/\gamma$ by Alice; the time for that distance to move past her will be $(L - 2\ell)/\gamma v_{0}$: the coasting time as measured by Alice is exactly what Bill ascribed to her. From Bill's point of view, Alice's clock is time dilated; from Alice's point of view, the distance is Lorentz contracted. But as the speed is the same for both twins, the coasting time Bill believes Alice experiences, and the time Alice measures for that part of the $S$ frame to coast past her, are the same. The time Alice measures for the $S$ frame to accelerate from $v = 0$ to $v= v_{0}$ is just Alice's proper time $\tau$, which was the time $\tau_{A}$ in Bill's measurement (\ref{eq:Alice_proper_time_accel}). The total acceleration time is just four times this. That is,
\begin{equation}\label{eq:t_Alice_Alice}
\begin{aligned}
\Delta t_{\text{Alice}} &= t_{\text{coast, Alice}} + t_{\text{accel, Alice}}\\
&= \dfrac{2L}{\gamma v_{0}} - \dfrac{4\ell}{\gamma v_{0}} + \dfrac{4c}{g}\text{arc sinh}(\gamma v_{0}/c)\\
&= \Delta t_{\text{Alice}}' 
\end{aligned}
\end{equation}
There is complete agreement between the travel time Bill measures on Alice's clock, and the travel time Alice measures on Alice's clock, as in all previous views.

What time does Alice think Bill measures? Special relativity requires that during the coasting, Alice should measure on Bill's clock $(1/\gamma)$ of the interval she measures on her own clock. Call this time $t_{\text{coast}}'$; 
\begin{equation}\label{eq:def_t_prime_coast}
t_{\text{coast}}' = \dfrac{t_{\text{coast, Alice}}}{\gamma} = \Bigl(\dfrac{1}{\gamma}\Bigr)\dfrac{2(L-2\ell)}{\gamma v_{0}} = \dfrac{2(L-2\ell)}{\gamma^{2}v_{0}}
\end{equation}
\indent The hard part is to calculate the acceleration times that Alice ascribes to Bill's clock, using the M\o ller coordinate for $X(T)$ (\ref{eq:X(T)}). The three periods of acceleration of $S$ take place at two different locations with respect to Alice; the first and last, when Alice is near Earth; and the second, when Alice is near Proxima. Let $t_{1}'$ be the time interval measured by Alice on Bill's clock during the first acceleration, near the Earth. By symmetry, this will equal the time interval of the last acceleration, as Alice brakes to arrive back on Earth. With the choices $T_{0} = 0$ and  $X(T_{0}) = 0$, $K = 1$. That is, near the Earth, 
\begin{equation}\label{eq:X(T)_E}
\begin{aligned}
X_{E}(T) &= (c^{2}/g) (\mathrm{sech}(gT/c) - 1))\\
V_{E}(T) &= -c\,\mathrm{sech}(gT/c)\,\tanh(gT/c)
\end{aligned}
\end{equation}
Note that the velocity at $T > 0$ is negative, indicating the Bill-Proxima frame moves to Alice's left. The time $t_{1}'$ ascribed by Alice to Bill's clock is then ($\tau_{A}$ is the proper time $\tau$ (\ref{eq:Alice_proper_time_accel}) Alice measures on her own clock during the acceleration) 
\begin{equation}\label{eq:t_1_prime}
\begin{aligned}
t_{1}' &=\int d\tau = \int_{0}^{\tau_{A}} \sqrt{\Bigl(1 + \dfrac{gX_{E}}{c^{2}}\Bigr)^{\!2} - \Bigl(\dfrac{V_{E}}{c}\Bigr)^{\!2}}\,dT\\
 &= \int_{0}^{\tau_{A}} \mathrm{sech}^{2}(gT/c)\,dT = \dfrac{c}{g}\tanh (g\tau_{A}/c) = \dfrac{v_{0}}{g}
\end{aligned}
\end{equation}
the last equality following from (\ref{eq:v_acc(tau)}) or from (\ref{eq:Alice_proper_time_accel}).

Similarly, let $t_{2}'$ be the time as measured by Alice on Bill's clock for $S$ to speed up from $v = 0$ to $v =  v_{0}$ near the star. By symmetry, Alice observes Bill's clock to record $2t_{2}'$ during the period of Proxima's acceleration near her. The function $X(T)$ must again be fit to the appropriate boundary conditions. However, the direction of the acceleration is now reversed, and so
\begin{equation}\label{eq:X(T)_P_prelim}
\begin{aligned}
X_{P}(T) &= (c^{2}/(-g)) (K\mathrm{sech}((-g/c)(T - T_{0})) - 1))\\
&= -(c^{2}/g)(K\mathrm{sech}((g/c)(T - T_{0})) - 1)) 
\end{aligned}
\end{equation}
because $\mathrm{sech}(x)$ is an even function. The moment that the Proxima clock reaches Alice, the velocity of every frame is zero, Bill's position is $-L$ according to Alice, and the time on Alice's clock is $\tfrac{1}{2}\Delta t_{\text{Alice}}$. Reset the clock, calling this value $T_{0}$, and calculate the time as measured by Alice for Bill to travel from $-L$ to $-L + \ell$ (or equivalently, for Proxima to travel from its closest approach to a distance $\ell$ away). At $T = \tfrac{1}{2}\Delta t_{\text{Alice}}$, the boundary condition says
\begin{equation}
\begin{aligned}
X_{P}(0) &= -(c^{2}/g) \left[K\,\mbox{sech}(0) - 1\right] = - L;\\
K &= 1 + (gL/c^{2})
\end{aligned}
\end{equation}
so that, letting $w = (g/c)(T - \tfrac{1}{2}\Delta t_{\text{Alice}})$,
\begin{equation}\label{eq:X(T)_P}
\begin{aligned}
X_{P}(T) &= - (c^{2}/g) \left[\Bigl(1 + \dfrac{gL}{c^{2}}\Bigr)\,\mbox{sech}\,w - 1\right]\\
V_{P}(T) &= c\Bigl(1 + \dfrac{gL}{c^{2}}\Bigr)\tanh w\,\mbox{sech}\,w
\end{aligned}
\end{equation}
Note $V_{P}(T) > 0$ for $T > \tfrac{1}{2}\Delta t_{\text{Alice}}$, indicating motion to Alice's right. The factor $(1 + gX/c^{2})$ becomes
\begin{equation}
1 + \dfrac{(-g)X_{P}}{c^{2}} = \Bigl(1 + \dfrac{gL}{c^{2}}\Bigr)\,\mbox{sech}\,w
\end{equation}
Then from (\ref{eq:t_1_prime})
\begin{equation}\label{eq:turnaround_time_alice}
\begin{aligned}
t_{2}' &= \int_{\tfrac{1}{2}\Delta t_{\text{Alice}}}^{\tfrac{1}{2}\Delta t_{\text{Alice}} + \tau_{A}}\sqrt{\Bigl(1 + \dfrac{gX_{P}}{c^{2}}\Bigr)^{\!2} - \Bigl(\dfrac{V_{P}}{c}\Bigr)^{\!2}}\,dT\\
&=\Bigl(1 + \dfrac{gL}{c^{2}}\Bigr)t_{1}'  =  \Bigl(1 + \dfrac{gL}{c^{2}}\Bigr)\dfrac{v_{0}}{g}
\end{aligned}
\end{equation}
For large $L$, the time $2t_{2}'$ is the dominant contribution (with this itinerary, 8 years) to $\Delta t'_{\text{Bill}}$. It arises from the large gravitational potential difference between the twins when they are at their greatest separation. The difference $2(t_{2}' - t_{1}')$ is the same quantity (\ref{eq:Born_term}) Born and Tolman derived from the equivalence principle, though their solution was approximate; they dropped the two acceleration times $(v_{0}/g)$ near the Earth. The total time $\Delta t'_{\text{Bill}}$ is given by \begin{equation}\label{eq:t_Bill_Alice}
\begin{aligned}
\Delta t'_{\text{Bill}} &= t'_{\text{coast}} + t'_{\text{accel}} = \dfrac{2(L - 2\ell)}{\gamma^{2}v_{0}} + 2t_{1}' + 2t_{2}'\\
&= \dfrac{2(L - 2\ell)}{\gamma^{2}v_{0}} + \dfrac{2v_{0}}{g} + \Bigl(1 + \dfrac{gL}{c^{2}}\Bigr)\dfrac{2v_{0}}{g}\\
&= \dfrac{2(L - 2\ell)}{\gamma^{2}v_{0}} + \dfrac{4v_{0}}{g} + \dfrac{2v_{0}L}{c^{2}}
\end{aligned}
\end{equation}
This isn't obviously the same as (\ref{eq:t_Bill}) as it must be to reconcile the clocks. The first term is, using (\ref{eq:ell}), 
\begin{equation}\label{eq:t'_coast_Bill}
\begin{aligned}
\dfrac{2(L - 2\ell)}{\gamma^{2}v_{0}} &= \dfrac{2L}{v_{0}}\Bigl(1 - \dfrac{v_{0}^{2}}{c^{2}}\Bigr) - \dfrac{4\ell}{v_{0}} \bigl(1 - \dfrac{v_{0}^{2}}{c^{2}}\big)\\
&=  \dfrac{2L}{v_{0}}\Bigl(1 - \dfrac{v_{0}^{2}}{c^{2}}\Bigr) - \dfrac{4\ell}{v_{0}} +  \dfrac{4v_{0}}{g}\left[\gamma - 1\right]\\
&=  \dfrac{2L}{v_{0}} - \dfrac{4\ell}{v_{0}} - \dfrac{2v_{0}L}{c^{2}} + \dfrac{4\gamma v_{0}}{g} - \dfrac{4v_{0}}{g}
\end{aligned}
\end{equation}
Substituting (\ref{eq:t'_coast_Bill}) into (\ref{eq:t_Bill_Alice}), one obtains finally
\begin{equation}\label{eq:t_Bill_Alice_final}
\begin{aligned}
\Delta t'_{\text{Bill}} &= \dfrac{2L}{v_{0}} - \dfrac{4\ell}{v_{0}} + \Bigl[\dfrac{2v_{0}L}{c^{2}} + \dfrac{4v_{0}}{g}\Bigr](1-1) + \dfrac{4\gamma v_{0}}{g} \\
&= \dfrac{2L}{v_{0}} - \dfrac{4\ell}{v_{0}} + \dfrac{4\gamma v_{0}}{g}\\
&= \Delta t_{\text{Bill}} 
\end{aligned}\tag{80}
\end{equation}
Bill's time as measured by Alice is \emph{exactly the same} as the time measured by Bill himself, $\Delta t_{\text{Bill}}$. General relativity is not required, and special relativity can provide an exact resolution, contrary to what is commonly found in the literature.

The ``paradox'' is resolved. Not merely do the reunited twins agree as to who is the elder, they agree on their different ages. Moreover, for this simple itinerary, the agreement holds good regardless of the value of the acceleration $g$, so long as it is nonzero. Presumably the clocks can be reconciled for \emph{any} itinerary. As M\o ller showed eighty years ago, (notes \ref{fn:Moller_Lorentz}; \ref{fn:Moller_Rel}, p.\,296), in the limit of infinite acceleration, these times---$\Delta t_{\text{Bill}}$, (\ref{eq:t_Bill}) and $\Delta t_{\text{Alice}}$, (\ref{eq:t_Alice_Alice})---agree with the approximate values (\ref{eq:approx_times});
\begin{equation}
\lim_{g \to \infty} \Delta t_{\text{Bill}} = \dfrac{2L}{v_{0}}\,; \qquad \lim_{g \to \infty} \Delta t_{\text{Alice}} = \dfrac{2L}{\gamma v_{0}}.
\end{equation}
Graphs for the worldlines for Alice's journey (her position as measured in the $S$ frame) in terms of Bill's proper time $t$ and Alice's proper time $\tau$, and for Bill's proper time $t'$ according to Alice in terms of the M\o ller coordinates, are given in Fig.\,\ref{fig:worldlines}. The functions shown in the graphs are discussed in the Appendix.

To conclude, there are full general relativistic solutions to the twins, but as shown here, there's no need to invoke general relativity. Special relativity's ability to handle accelerated frames, and the equivalence principle, provide exact and consistent answers at the twins' reunion. \\

\begin{figure}[!htbp]
\centering
\scalebox{0.625}{\includegraphics{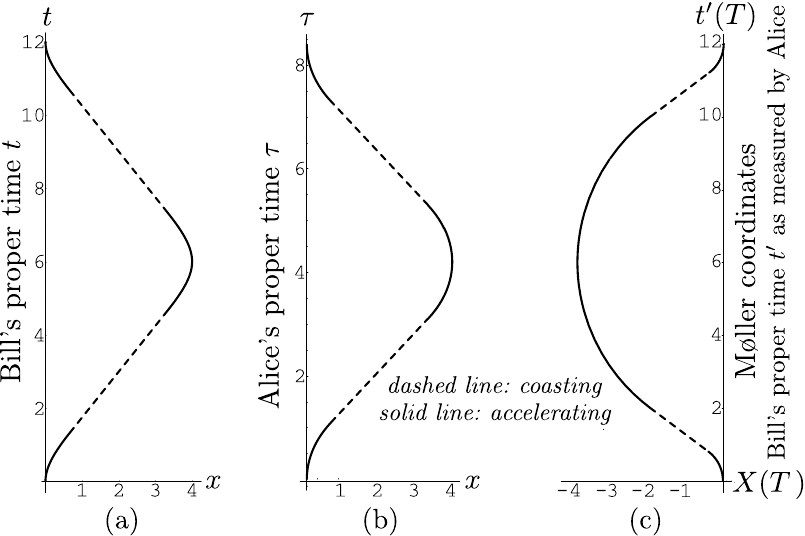}}
\caption{Worldlines of Alice's journey}
\label{fig:worldlines}
\end{figure}

 \normalsize

\small

\begin{appendices}

\section*{\emph{Lagniappe}: Feynman's puzzle}

In his memoir,\footnote{\emph{Surely You're Joking, Mr.\ Feynman!}, Richard P.\ Feynman (as told to Ralph Leighton), W.\,W.\,Norton, New York, 1985, p.\,37; \emph{cf.}\ \emph{The Feynman Lectures on Gravitation}, R.\,P.\,Feynman, F.\,B.\,Moringo, W.\,G.\,Wagner, ed.\,Brian Hatfield, Penguin Books, London, 1999, ``Maximum clock rates in gravity fields,'' pp.\,95-97.} Richard Feynman describes a puzzle he posed to an assistant of Einstein's. 
\begin{quote}
You blast off in a rocket which has a clock on board, and there's a clock on the ground. The idea is that you have to be back  when the clock on the ground says that one hour has passed. Now you want it so that when you come back, your clock is as far ahead as possible. According to Einstein, if you go very high, your clock will go faster, because the higher something is in a gravitational field, the faster its clock goes. But if you try to go too high, since you've only got an hour, you have to go so fast to get there that the speed slows your clock down. So you can't go too high. The question is, exactly what program of speed and height should you make so that you get the maximum time on your clock?
\end{quote}
The assistant might have seen the answer at once (but did not). It is a principle in relativity that free fall maximizes proper time, and so that is the answer: the rocket is simply given an initial velocity to reach a maximum height and fall back in the time specified (as will be seen, an hour is wildly impractical). The motion of the rocket is formally identical to Alice's motion at her turnaround, which is why the puzzle is included here.

A quarter century later Feynman restated this puzzle and provided the solution in a lecture.\footnote{\label{fn:Hughes}\emph{Notes from Richard Feynman's Hughes Lectures 1966-1971}, ed.\ John T. Neer, \url{https://thehugheslectures.info}, vol.\,1, pp.\,30-40.} Here is an excerpt of his solution.

\begin{quote}
``The traveler observes two moments on his clock separated by a time $d\tau$. We relate this time to the proper time on earth, $dt$. Since the guy is higher, he looks purple because of gravity, so 
\[
d\tau = dt\left(1 + \frac{\varphi}{c^{2}}\right) \tag{F1}
\]
Also because he is moving at a good clip, the time dilation arising from our approximation is 
\[
d\tau = dt - \frac{v^{2}}{2c^{2}}\,dt \tag{F2}
\]
so the proper time on earth for the interval is
\[
d\tau = dt + \frac{\varphi}{c^{2}}\,dt - \frac{v^{2}}{2c^{2}}\,dt \tag{F3}
\]
integrated over $0 \le t \le T$\dots The time interval seen on the [traveler's] wrist watch is then 
\[
\begin{aligned}
\tau_{2} - \tau_{1} &= \int_{0}^{T} \left[1 + \frac{\varphi}{c^{2}} - \frac{v^{2}}{2c^{2}}\right]\,dt\rule[-2ex]{0pt}{2ex}\\
&= T + \frac{1}{c^{2}} \int_{0}^{T} \left[\varphi(x(t))- \frac{\dot{x}^{2}}{2}\right]\,dt
\end{aligned} \tag{F4}
\]
so what we want is to make the bracket term an extremum. Since $\varphi = gx(t)$, we have 
\[\label{eq:Feynman_approx_int}
\Delta\tau = T +  \frac{1}{c^{2}} \int_{0}^{T} \left[gx(t) - \frac{\dot{x}^{2}}{2}\right]\,dt \tag{F5}
\]
We can solve this problem by the principle of least action.''
\end{quote}
Feynman finds the extremum not with the Euler-Lagrange equation, but with his preferred approach,\footnote{Richard P.\,Feynman, Robert B.\,Leighton, Matthew Sands, \emph{The Feynman Lectures on Physics}, Addison-Wesley, Reading, MA, 1964, vol.\,II, pp.\,19-4--19-5.} and obtains $g = - \ddot{x}$, ``just as it would be for a free falling body.'' He concludes
\begin{quote} 
``The consequence of this little exercise was very important to Einstein, who seized it and concluded that since local clocks always measure the longest time, clock [rates] will vary under different circumstances; i.e., there is no absolute time. The rate of time progression depends on where you are in space and whose clock you use.''
\end{quote}
In the lecture, Feynman asked for the ground and wristwatch times, but did not provide them. Calling the ground time $T$, using the Galilean forms for $v$ and $x$ to solve the approximate equation $\ddot{x} = - g$, 
\[
v = v_{0} - gt, \qquad x = v_{0}t - \tfrac{1}{2}gt^{2}
\]
with $\varphi = gx$, $v_{0} = gt_{h} = \tfrac{1}{2}gT$, and $h = v_{0}^{2}/2g$,
\begin{equation}\label{eq:Feynman_t}
\begin{aligned}
\Delta\tau &= 2\int_{0}^{\tfrac{1}{2}T}\left[1 - \tfrac{1}{2}\dfrac{v^{2}}{c^{2}} + \dfrac{gx}{c^{2}}\right]\,dt\\
&= T\Bigl(1 + \tfrac{1}{6}\dfrac{v_{0}^{2}}{c^{2}}\Bigr)
\end{aligned}
\end{equation}
\indent That's the approximate, Galilean answer. Feynman then describes Einstein's altering the Minkowski metric to the form\footnote{Note \ref{fn:Hughes}, p.\,40. The exponent of the coefficient of $dt^{2}$ is erroneously given as 1.}
\[
ds^{2} = \left(1 + \frac{\varphi}{c^{2}}\right)^{\!2}\!dt^{2} - \alpha\,dx^{2} - \beta\,dy^{2} - \gamma\,dz^{2} \tag{F6}
\]
suggesting that the exact answer would be found using
\begin{equation}
ds^{2} =  c^{2}\left[\left(1 + \frac{\varphi}{c^{2}}\right)^{\!2} - \frac{v^{2}}{c^{2}}\right]dt^{2} = c^{2}d\tau^{2}
\end{equation}
(with $dy = dz = 0$) because, for $\varphi \ll c^{2}$ and $v \ll c$,
\begin{equation}
d\tau \approx \sqrt{1 + \frac{2\varphi}{c^{2}} - \frac{v^{2}}{c^{2}}} \approx 1 + \frac{\varphi}{c^{2}} - \frac{v^{2}}{2c^{2}} 
\end{equation}
as he used for his integral, (\ref{eq:Feynman_approx_int}). The quantity to be maximized for the exact answer becomes
\begin{equation}
\int_{0}^{T} \sqrt{\left(1 + \frac{gx}{c^{2}}\right)^{2} - \frac{v^{2}}{c^{2}}}\,dt
\end{equation}
precisely equivalent to (\ref{eq:proper_time_int}). From (\ref{eq:X(T)}) the solution is
\begin{equation}\label{eq:Feynman_x}
x(t) = (c^{2}/g)[K\mbox{sech}((g/c)(t - t_{0})) - 1]
\end{equation}
and taking the derivative,
\begin{equation}\label{eq:Feynman_v}
v(t) = - cK\tanh((g/c)(t - t_{0}))\,\mbox{sech}((g/c)(t - t_{0}))
\end{equation}
As with Alice's turnaround, choose $t_{0}$ to be the time $t_{h} = \tfrac{1}{2}T$ when the rocket is at its greatest distance from the Earth; let $h$ be this distance as measured by the person on the ground. To find $K$ and $v_{0}$, set $t = 0$ when $x = 0$:
\begin{equation}
\begin{aligned}
K &= \cosh(gt_{h}/c);\\
v(0) &\equiv v_{0} =  c\tanh(gt_{h}/c)
\end{aligned}
\end{equation}
Additionally,
\begin{equation}\label{eq:Feynman_gamma}
\gamma_{0} = \dfrac{1}{\sqrt{1 - (v_{0}/c)^{2}}} = \cosh(gt_{h}/c) = K
\end{equation}
which agrees with (\ref{eq:gamma_acc(tau)}) at $t= 0$. Using (\ref{eq:Feynman_x}) and setting $x(t _{h}) = h$,
\begin{equation}\label{eq:Feynman_h}
h = (c^{2}/g)[\gamma_{0} - 1]
\end{equation}
in accord with (\ref{eq:position_accel}). Letting $\Delta\tau$ be the travel time on the rocket clock,
\begin{equation}
\Delta\tau = 2\int_{0}^{t_{h}}\sqrt{\left(1 + \frac{gx}{c^{2}}\right)^{2} \!- \frac{v^{2}}{c^{2}}}\,dt
\end{equation}
The integral has already been calculated in (\ref{eq:turnaround_time_alice}); one need only replace $L$ by $h$:
\begin{equation}
\begin{aligned}
\Delta\tau &= \dfrac{2v_{0}}{g}\Bigl(1 + \dfrac{gh}{c^{2}}\Bigr) = T \Bigl(1 + \dfrac{gh}{c^{2}}\Bigr)\\
 &= \gamma_{0}T \approx T\Bigl(1 + \tfrac{1}{2}\dfrac{v_{0}^{2}}{c^{2}}\Bigr)
\end{aligned}
\end{equation}
using (\ref{eq:Feynman_h}) for $h$. The free fall rocket (proper) time is simply $\gamma_{0}$ times the ground (proper) time, as might have been guessed, and a little larger than the Galilean value (\ref{eq:Feynman_t}).

Does the exact solution, like the approximate, describe the position as quadratic in time, and the velocity as linear, at least for $v\ll c$? Expanding (\ref{eq:Feynman_v}) to second order in $(v/c)$, with $v_{0} = gt_{h}$, one finds
\begin{equation}
\begin{aligned}
v &= (v_{0} - gt)\Bigl(1 + \frac{v_{0}^{2}}{c^{2}}\Bigr)\\
x(t) &= gt_{h}t - \tfrac{1}{2}gt^{2} = v_{0}t - \tfrac{1}{2}gt^{2}
\end{aligned}
\end{equation}
The position is quadratic in time; the velocity, while linear, is not exactly the derivative of the position; see the graphs of the exact functions (\ref{eq:Feynman_x}) and (\ref{eq:Feynman_v}). For larger values of $v$, the graphs diverge more from the Galilean forms.

Feynman's analysis assumes a constant value of $g$ and an hour travel time for $T$. These two conditions cannot at the same time be satisfied on Earth. During its return, half an hour in duration absent air friction, the rocket would fall under a constant $g \sim 10$ m/s\textsuperscript{2} about 2.5 Earth radii; but at $h = 2.5 R_{E}$, the value of $g$ is less than a tenth of its ground value: $g$ is not constant. To bring out differences between the classical and the relativistic answers, $v_{0}$ should be large enough to produce deviations from Galilean results. What is wanted is a large $g$ which does not vary much over a great distance, as found outside the event horizon of a black hole. Suitable values are, with $t_{h} = 1800$\,s,
\begin{equation}
v_{0} = \tfrac{7}{25}c = 8.4 \times 10^{7}\,\mathrm{m/s}; \quad \gamma_{0} = \tfrac{25}{24};
\end{equation}
which from (\ref{eq:Feynman_gamma}) leads to\footnote{The Galilean value $g_{\textrm{Gal}}$ equals $v_{0}/t_{h}$, or $4.67 \times 10^{4}\,\mathrm{m/s}^{2}$.}
\begin{equation}
g _{0}= \dfrac{c}{t_{h}}\cosh^{-1}\gamma =  4.79 \times 10^{4}\,\mathrm{m/s}^{2} = 9.58 \times 10^{-3}\,\dfrac{c}{\text{min}}
\end{equation}
and from (\ref{eq:Feynman_h})
\begin{equation}
h = (c^{2}/g_{0})[\gamma_{0} - 1] = 7.82 \times 10^{10}\,\mathrm{m} = 4.34\, c\text{-min}
\end{equation}
It's impossible to have $g$ constant over this distance, so try to find starting and ending distances $R_{1}, R_{2}$ such that the average of $g(R_{1}) = g_{1}$ and $g(R_{2}) = g_{2}$ equals $g_{0}$, with $R_{2} - R_{1} = h$. One way to choose the black hole mass is to ensure that $R_{1}$ is greater than the black hole's Schwarzschild radius $R_{S}$. Let $R_{0}$ be the distance from the center of the black hole such that the local $g$ there equals $g_{0}$, and let $R_{1} = R_{0} - \tfrac{1}{2}h$. The condition $R_{1} = R_{S}$ is stated
\begin{equation}
\sqrt{GM}{g_{0}} - \tfrac{1}{2}h = \dfrac{2GM}{c^{2}}
\end{equation}
which is a quadratic in $\sqrt{GM}$. The smaller value of $M$ leads to a greatly varying $g$, so choose the larger (rounding up):
\begin{equation}
M = 5.78 \times 10^{38}\,\mathrm{m}
\end{equation}
(the black hole at the center of the Milky Way, Sagittarius A*, has a mass of about $8 \times 10^{36}$ kg; NGC 6166 has a mass of approximately $5.8 \times 10^{40}$ kg). The various $R$ values for this mass are
\begin{equation}
\begin{gathered}
R_{S} = \dfrac{2GM}{c^{2}} = 8.57 \times 10^{11}\,\text{m}\\
R_{0} = \sqrt{\dfrac{GM}{g_{0}}} = 8.97 \times 10^{11}\,\text{m}\\
R_{1} = R_{0} - \tfrac{1}{2}h = 8.58 \times 10^{11}\,\text{m}\\
R_{2} =  R_{0} + \tfrac{1}{2}h = 9.36 \times 10^{11}\,\text{m}
\end{gathered}
\end{equation}
(note that all the $R_{i}$ are greater than $R_{S}$). The gravitational fields are
\begin{equation}
g_{1} = 5.24 \times 10^{4}\,\mathrm{m/s}^{2}; \quad g_{2} = 4.40 \times 10^{4}\,\mathrm{m/s}^{2}
\end{equation}
and the average of these values is within 1\% of $g_{0}$. These values (nearly) meet Feynman's conditions, and the new condition that $v_{0}$ is an appreciable fraction of the speed of light. 

Below are graphs of the position (in light-minutes), velocity (in fractions of $c$) and time (in minutes). Note that $h_{\text{Gal}} = 4.2$ $c$-min, while $h_{\text{rel}} = 4.34$ $c$-min. Even at $v/c = \tfrac{7}{25}$, the differences are small.
\begin{figure}[!htbp]
\centering
\scalebox{0.7}{\includegraphics{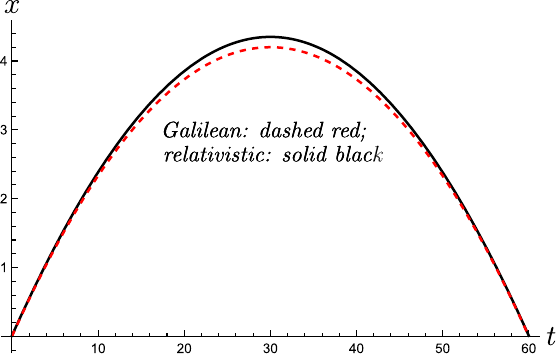}}
\caption{Positions of vertically launched rocket, $v_{0} = \tfrac{7}{25}c$}
\label{fig:x_rocket}
\end{figure}
\vspace*{-0.25in}

\begin{figure}[!htbp]
\centering
\scalebox{0.7}{\includegraphics{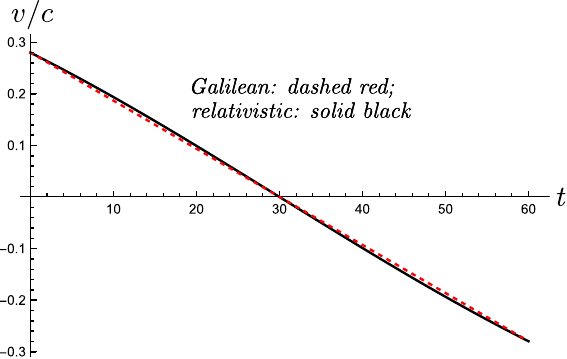}}
\caption{Velocities of vertically launched rocket, $v_{0} = \tfrac{7}{25}c$}
\label{fig:v_rocket}
\end{figure}


\section*{Appendix: The worldline graphs in detail}

Alice's itinerary is symmetric about the turnaround point in space and time. Each of the worldline graphs are thus symmetric about reflection through the horizontal line which intersects a given time axis at the turnaround time, and it suffices to consider only the first half of the journeys. 

Consider Alice's position according to Bill's coordinate system (Fig.\,\ref{fig:worldlines} (a)). During the first acceleration, Bill observes Alice's position $x$ as determined by (\ref{eq:position_time_Bill});
\begin{equation}\label{eq:bill_alice_acc_1}
x = (c^{2}/g)\bigl[\sqrt{1 + (gt/c)^{2}} - 1\bigr], \quad 0\le t \le t_{B} = \gamma v_{0}/g
\end{equation}
In keeping with relativistic conventions, the graphs plot $t$ on the abcissa. Inverting this relation, ($\ell = (c^{2}/g)[\gamma - 1]$, (\ref{eq:ell}))
\begin{equation}
t = (c/g)\sqrt{(1 + [gx/c^{2}])^{2} - 1}, \quad 0 \le x \le \ell 
\end{equation}
Bill describes the first period of Alice's coasting by the line
\begin{equation}\label{eq:bill_alice_coast_1}
x - \ell = v_{0}(t - t_{B}), \quad t_{B} < t \le \tfrac{1}{2}\Delta t_{\text{Bill}} - t_{B}
\end{equation} 
Let $x_{c}$ be the position (as measured by Bill) at the end of coasting; $x_{c} = L - \ell$. Then
\begin{equation}\label{eq:bill_alice_coast_1_t}
t = (1/v_{0})(x - \ell) + t_{B}, \quad \ell < x \le x_{c}
\end{equation}
The graphs of the two periods must be continuous but also smooth: the derivatives must match at $t = t_{B}$. For the first acceleration,
\begin{equation}
\left.\dfrac{dx}{dt}\right|_{t_{B}} = \dfrac{gt_{B}}{\sqrt{1 + (gt_{B}/c)^{2}}} = v_{0}
\end{equation}
which is equal to the derivative of the line (\ref{eq:bill_alice_coast_1}) at $t_{B}$ (and at all other times).

The first half of Alice's second acceleration may conveniently be described by 
\begin{equation}
\begin{aligned}
x &= L - (c^{2}/g)\bigl(\sqrt{1 + \bigl[g(t - \tfrac{1}{2}\Delta t_{\text{Bill}})/c\bigr]^{2}} - 1\bigr),\\
&\tfrac{1}{2}\Delta t_{\text{Bill}} - t_{B} < t \le \tfrac{1}{2}\Delta t_{\text{Bill}} 
\end{aligned}
\end{equation}
(note that the sign of $g$ has been changed) so that $t = \tfrac{1}{2}\Delta t_{\text{Bill}}$ when $x = L$. This is inverted to give
\begin{equation}\label{eq:bill_alice_sec_accel}
\begin{aligned}
t &= \tfrac{1}{2}\Delta t_{\text{Bill}} - (c/g)\sqrt{\bigl[1 + g(L-x)/c^{2}\bigr]^{2} - 1},\\
&x_{c} < x \le L
\end{aligned}
\end{equation}
The derivatives of the line and the curve at $t = \tfrac{1}{2}\Delta t_{\text{Bill}} - t_{B}$ are equal;
\begin{equation}
\left.\dfrac{dx}{dt}\right|_{\tfrac{1}{2}\Delta t_{\text{Bill}} - t_{B}} = - \dfrac{g(-t_{B})}{\sqrt{1 + (g(-t_{B})/c)^{2}}} = v_{0}
\end{equation}
as required. The function $t(x)$ can then be described by
\begin{equation}\label{eq:Bill_Alice_out}
t = \begin{cases}
         \Bigl(\dfrac{c}{g}\Bigr)\sqrt{\left[1 + \dfrac{gx}{c^{2}}\right]^{2} - 1}, &0 \le x \le \ell \rule[-3.75ex]{0pt}{3.75ex}\\
         \dfrac{1}{v_{0}}(x - \ell) + t_{B},  & \ell < x \le x_{c}\rule[-2ex]{0pt}{2ex}\\
         \tfrac{1}{2}\Delta t_{\text{Bill}} - \Bigl(\dfrac{c}{g}\Bigr)\sqrt{\left[1 + \dfrac{g(L-x)}{c^{2}}\right]^{2} \!\!- 1}, &x_{c} < x \le L
        \end{cases}
\end{equation}
This function only describes the trip out. The function for the return leg of the trip is obtained by replacing $t$ by $\Delta t_{\text{Bill}} - t$, and solving for $t$.        

The second graph (Fig.\,\ref{fig:worldlines} (b)) is Alice's world line according to her own clock (her proper time), but in Bill's coordinate frame. Her first acceleration is described by inverting (\ref{eq:position_accel});
\begin{equation}\label{eq:alice_first_accel_tau}
\tau = (g/c)\,\mathrm{arc cosh}\,\bigl[1 + (gx/c^{2})\bigr], \quad 0 \le x \le \ell
\end{equation}
Note that at the upper limit of this expression, ($\ell = (c^{2}/g)[\gamma - 1]$)
\begin{equation}
\begin{aligned}
\tau &= (c/g)\,\text{arc cosh}[1 + (g\ell/c^{2})] = (c/g)\,\text{arc cosh}(\gamma)\\ 
&= (c/g)\,\text{arc sinh}(\gamma v_{0}/c) = (c/g)\,\text{arc sinh}(t_{B}) = \tau_{A}
\end{aligned}
\end{equation}
as in (\ref{eq:Alice_proper_time_accel}). Describe the first coasting period by the same line as before, (\ref{eq:bill_alice_coast_1}), as both Alice and Bill are in Lorentz-related inertial frames during this part of the journey. Because of time dilation, 
\begin{equation}
t - t_{B} = \gamma(\tau - \tau_{A}) 
\end{equation}
where $\tau$ and $\tau_{A}$ are Alice's proper times corresponding to Bill's times $t$, $t_{B}$, respectively. That is, the relevant line becomes, in place of (\ref{eq:bill_alice_coast_1}), the line
\begin{equation}\label{eq:alice_first_coast_x}
x - \ell = \gamma v_{0}(\tau - \tau_{A})
\end{equation}
or in other words,
\begin{equation}\label{eq:alice_first_coast_tau}
\tau = (1/\gamma v_{0})(x - \ell) + \tau_{A}, \quad \ell < x \le L - \ell
\end{equation}
Do the derivatives of the functions match at $\tau = \tau_{A}$? By inspection, the constant slope of the line (\ref{eq:alice_first_coast_x}) at $\tau = \tau_{A}$ is $\gamma v_{0}$. The derivative of the position (\ref{eq:position_accel}) at that instant is
\begin{equation}\label{eq:deriv_alice_acc}
\begin{aligned}
\left.\dfrac{dx}{d\tau}\right|_{\tau_{A}} &= \dfrac{d}{d\tau} (c^{2}/g)\bigl[\cosh(g\tau/c) - 1\bigr] = c\sinh(g\tau_{A}/c)\\
 &= c\sinh(\text{arc sinh}\,(\gamma v_{0}/c)) = \gamma v_{0}
\end{aligned}
\end{equation}
in agreement with the line. 

For Alice's second acceleration, return to (\ref{eq:position_accel}) and fit it to the new circumstances. Change the sign of $g$, and ensure that at $\tau = \tfrac{1}{2}\Delta t_{\text{Alice}}$, Alice is at $x = L$:
\begin{equation}\label{eq:bill_alice_accel_2}
x = L - (c^{2}/g)\bigl[\cosh(g/c)(\tfrac{1}{2}\Delta t_{\text{Alice}} - \tau) - 1\bigr]
\end{equation}
which when inverted becomes
\begin{equation}\label{eq:bill_alice_accel_2_tau}
\begin{aligned}
\tau &= \tfrac{1}{2}\Delta t_{\text{Alice}} - (c/g)\,\text{arc cosh}\,\bigl[1 + g(L-x)/c^{2}\bigr],\\
&L- \ell < x \le L
\end{aligned}
\end{equation}
The derivative of $x$ as specified in (\ref{eq:bill_alice_accel_2}) at $\tau = \tfrac{1}{2}\Delta t_{\text{Alice}} - \tau_{A}$, when the second acceleration begins, is also $\gamma v_{0}$, so the line and the curve of the second acceleration join smoothly. Then $\tau(x)$ is given by ($x_{c} = L - \gamma$)
\begin{equation}\label{eq:Alice_Alice_out}
\tau = \begin{cases}
\Bigl(\dfrac{c}{g}\Bigr)\,\mathrm{arc\,cosh}\,\Bigl(1 + \dfrac{gx}{c^{2}}\Bigr), &\!\!\!0 \le x \le \ell\rule[-3.75ex]{0pt}{3.75ex}\\
\Bigl(\dfrac{1}{\gamma v_{0}}\Bigr)(x - \ell) + \tau_{A}, &\!\!\!\ell < x \le x_{c}\rule[-2.5ex]{0pt}{2.5ex}\\
\tfrac{1}{2}\Delta t_{\text{Alice}} - \Bigl[\dfrac{c}{g}\Bigr]\mathrm{arc\,cosh}\Bigl[1 + \dfrac{g(L\!-\!x)}{c^{2}}\Bigr],&\!\!\!x_{c} < x \le L
\end{cases}
\end{equation}
As with the function (\ref{eq:Bill_Alice_out}) for $t(x)$, this function $\tau(x)$ only describes the trip out. The functions for the return leg of the trip are obtained by replacing $\tau$ by $\Delta t_{\text{Alice}} - \tau$, and solving for $\tau$.        

The third graph, Fig.\,\ref{fig:worldlines}\hspace*{2pt}(c), describes Bill's proper time $t'$ as Alice measures it in terms of her proper time $\tau = T$ and the coordinate $X$ that Alice uses to  locate Bill. The relevant M\o ller coordinates (\ref{eq:Moller_coords}) become
\begin{equation}\label{eq:alice_moller}
\begin{aligned}
t' &= \left[\dfrac{c}{g} + \dfrac{X}{c}\right] \sinh ((g/c)(T-T_{0})) + t'_{0}\rule[-3ex]{0pt}{3ex}\\
x' &= \left[\smash[t]{\dfrac{c^{2}}{g} }+ X\right] \cosh((g/c)(T-T_{0}) + x'_{0} \rule[-2ex]{0pt}{2ex}
\end{aligned}
\end{equation}
From the variational argument, one has the expression (\ref{eq:X(T)}) for $X(T)$. For the first acceleration, set $X(T)  = 0$ when $T = T_{0} = 0$, which makes $K = 1$, and
\begin{equation}\label{eq:alice_moller_acc_1}
X(T) = (c^{2}/g) \left[\mbox{sech} \left((g/c)(T - T_{0})\right) - 1\right]
\end{equation}
Substituting this form of $X(T)$ into $t'$ and $x'$ gives
\begin{equation}\label{eq:bill_moller}
\begin{aligned}
t' &= (c/g)\,\mbox{sech}\Bigl[\dfrac{g}{c}(T-T_{0})\Bigr]\sinh\Bigl[\dfrac{g}{c}(T-T_{0})\Bigr] + t'_{0}\\ 
&= (c/g)\tanh\Bigl[\dfrac{g}{c}(T-T_{0})\Bigr] + t'_{0}\rule[-2ex]{0pt}{2ex}\\
x' &=(c^{2}/g)\,\mbox{sech}\,\Bigl[\dfrac{g}{c}(T-T_{0})\Bigr]\cosh\Bigl[\dfrac{g}{c}(T-T_{0})\Bigr] + x'_{0}\\ 
&= (c^{2}/g) + x'_{0}
\end{aligned}
\end{equation}
With the choices $t' = 0$ and $x' = 0$ when $T = T_{0} = 0$, then $t'_{0} = 0$ and $x'_{0} = - (c^{2}/g)$. Making these choices,
\begin{equation}\label{eq:bill_moller_first_acc}
\begin{aligned}
t' &=(c/g)\tanh(gT/c)\\
x' &= 0\\
X &=  (c^{2}/g) \left[\mbox{sech}\,(gT/c) - 1\right]
\end{aligned}
\end{equation}
The equation $x' = 0$ simply says that from Bill's point of view, Bill doesn't move from his origin. Note also the formula for $t'$ is consistent with (\ref{eq:t_1_prime}). A graph of $t'$ in terms of $X$ becomes very easy to describe, owing to the identity $1 - \tanh^{2} u = \text{sech}^{2} u$. Solve for $\text{sech}$ and $\tanh$ and use the identity; obtain
\begin{equation}\label{eq:bill_moller_first_acc_graph}
t' = \dfrac{c}{g}\sqrt{1 - \Bigl(1 + \dfrac{gX}{c^2}\Bigr)^{2}}
\end{equation}
What interval describes $X$ during this acceleration? Find the final position $X_{1}$ of the acceleration with $t' = t_{B} = \gamma v_{0}/g$. From (\ref{eq:bill_moller_first_acc})
\begin{equation}
\begin{aligned}
T_{1} &= (c/g)\,\mathrm{arc\,tanh}\Bigl[\dfrac{\gamma v_{0}}{c}\Bigr] = (c/g)\,\mathrm{arc\,cosh}\,\gamma;\\
X_{1} &= (c^{2}/g)\!\bigl[\mbox{sech}\left(\mathrm{arc\,cosh}\,\gamma\right) -1\bigr] = (c^{2}/g)\left[\dfrac{1}{\gamma} - 1\right]\\
&= -\ell/\gamma
\end{aligned}
\end{equation}
Bill's first acceleration in the $X$-$t'$ variables is then given by
\begin{equation}\label{eq:bill_first_acc_graph}
t' = \dfrac{c}{g}\sqrt{1 - \Bigl(1 + \dfrac{gX}{c^2}\Bigr)^{2}}, \quad -\ell/\gamma \le X \le 0
\end{equation}
At the end of the acceleration, Bill's time is
\begin{equation}
t' = \dfrac{c}{g}\sqrt{1 - \Bigl(1 - \dfrac{g\ell}{\gamma c^2}\Bigr)^{2}} = \dfrac{v_{0}}{g} = t'_{1}
\end{equation}
consistent with (\ref{eq:t_1_prime}). For the first coasting period, the motion is described by a line of the form
\begin{equation}
X - X_{1} = \left.\dfrac{dX}{dt'}\right|_{t'_{1}}(t' - t'_{1}) 
\end{equation}
where, using (\ref{eq:def_t_prime_coast}),
\begin{equation}
\left(\dfrac{dX}{dt'}\right)^{-1} = \dfrac{dt'}{dX} = - \dfrac{1 + (gX/c^{2})}{c\sqrt{1 - (1 + (gX/c^{2})^{2}}}
\end{equation}
Putting $X = X_{1} = -\ell/\gamma$, $(1 + (gX/c^{2}) = 1/\gamma$, so
\begin{equation}
\left.\dfrac{dX}{dt'}\right|_{t'_{1}} = - \gamma v_{0}
\end{equation}
Then the line describing Bill's first coasting is
\begin{equation}
X + \ell/\gamma = - \gamma v_{0}(t' - t'_{1})
\end{equation}
Find $X_{c}$, the value of $X$, and $t'_{c}$, the time, at the end of Bill's first coasting by recalling (\ref{eq:def_t_prime_coast}):
\begin{equation}
\begin{aligned}
t'_{\text{coasting}} &= t'_{c} - t'_{1} = \dfrac{L - 2\ell}{\gamma^{2}v_{0}};\\
t'_{c} &=  \dfrac{L - 2\ell}{\gamma^{2}v_{0}} + \dfrac{v_{0}}{g}
\end{aligned}
\end{equation}
Then
\begin{equation}
\begin{gathered}
-\dfrac{X_{c} + (\ell/\gamma)}{\gamma v_{0}} = \dfrac{L - 2\ell}{\gamma^{2}v_{0}};\\
X_{c} = \dfrac{-L + \ell}{\gamma}
\end{gathered}
\end{equation}
The line for Bill's first coasting in terms of $X$ and $t'$ is then
\begin{equation}\label{eq:Alice_Bill_first_coast}
t' - t'_{1} = - \dfrac{1}{\gamma v_{0}}(X - X_{1}), \quad X_{c} \le X \le X_{1}
\end{equation}
There is a consistency check lurking inside this expression. During this period, the coordinates $(t', x')$ and $(\tau, X)$ are Lorentz-related, according to the usual rule:
\begin{equation}
\begin{aligned}
\tau &= \gamma \bigl(t' - \dfrac{v_{0}x'}{c^{2}}\bigr)\\
X &= \gamma(x' - v_{0}t')
\end{aligned}
\end{equation}
Substituting in for $X$, $X_{1}$ in (\ref{eq:Alice_Bill_first_coast}),
\begin{equation}
t' - t'_{1} = - \dfrac{1}{\gamma v_{0}}\left(\gamma(x' - x_{1}') - \gamma v_{0}(t' - t_{1}')\right)
\end{equation}
which implies
\begin{equation}
x' = x_{1}'
\end{equation}
That means as far as Bill is concerned, he is at the same place at the end of the first acceleration as he is at any time $t'$ during the coasting: he isn't moving. In particular, from (\ref{eq:bill_moller_first_acc}), he must be where he was at the beginning of the acceleration: the origin. 

For Bill's second acceleration, use the expression for $X$ in (\ref{eq:X(T)_P}), 
\begin{equation}\label{eq:bill_accel_sec_X}
\begin{aligned}
X &= - (c^{2}/g)\Bigl[\Bigl(1 + \dfrac{gL}{c^{2}}\Bigr)\,\mbox{sech}\,w - 1\Bigr], \\
w &= (g/c)(T - \tfrac{1}{2}\Delta t_{\text{Alice}})
\end{aligned}
\end{equation}
Using this expression for the M\o ller coordinate(\ref{eq:alice_moller}) for $t'$,
\begin{equation}\label{eq:bill_accel_sec_t_prime}
\begin{aligned}
t' &= -(c/g)\Bigl[1 - \dfrac{gX}{c^{2}}\Bigr]\sinh w + t_{0}'\\
&= -(c/g)\Bigl[1 + \dfrac{gL}{c^{2}}\Bigr]\tanh w + t_{0}'
\end{aligned}
\end{equation}
The value of $t_{0}'$ is determined by the condition that $t' = \tfrac{1}{2}\Delta t_{\text{Bill}}$ when $w = 0$ and $X = -L$; i.e. 
\begin{equation}
t'_{0} = \tfrac{1}{2}\Delta t_{\text{Bill}}
\end{equation}
Solve (\ref{eq:bill_accel_sec_X}) for $\text{sech}\,w$, and (\ref{eq:bill_accel_sec_t_prime}) for $\tanh w$, and use the trigonometric identity to obtain
\begin{equation}
\Bigl(\dfrac{g(t'-t'_{0})}{c}\Bigr)^{2} + \Bigl(1 - \dfrac{gX}{c^{2}}\Bigr)^{2} = \Bigl(1 + \dfrac{gL}{c^{2}}\Bigr)^{2}
\end{equation}
and consequently
\begin{equation}
t' - t'_{0} = \pm(c/g)\sqrt{\Bigl(1 + \dfrac{gL}{c^{2}}\Bigr)^{2} - \Bigl(1 - \dfrac{gX}{c^{2}}\Bigr)^{2}}
\end{equation}
For the interval $-L \le X < X_{c}$, $t' < t'_{0} = \tfrac{1}{2}\Delta t_{\text{Bill}}$, so 
\begin{equation}
t' = t'_{0} - (c/g)\sqrt{\Bigl(1 + \dfrac{gL}{c^{2}}\Bigr)^{2} - \Bigl(1 - \dfrac{gX}{c^{2}}\Bigr)^{2}}
\end{equation}
Check the smoothness at $t' = t_{c}'$ (or what is the same, at $X = X_{c}$). The derivative $dX/dt'$ of the line is $-\gamma v_{0}$, and
\begin{equation}
\left(\dfrac{dX}{dt'}\right)^{-1} = \dfrac{dt'}{dX} = - \dfrac{1 - (gX/c^{2})}{c\sqrt{\Bigl(1 + \dfrac{gL}{c^{2}}\Bigr)^{2} - \Bigl(1 - \dfrac{gX}{c^{2}}\Bigr)^{2}}
}
\end{equation}
Putting in $X_{c}$ for $X$ gives
\begin{equation}
1 - (gX_{c}/c^{2}) = 1 + \dfrac{g(L - \ell)}{\gamma c^{2}} = \dfrac{1}{\gamma}\left(1 + \dfrac{gL}{c^{2}}\right)
\end{equation}
and so
\begin{equation}
\left.\dfrac{dt'}{dX}\right|_{X_{c}} = - \dfrac{1}{c\gamma\sqrt{1 - (1/\gamma)^{2}}} = - \dfrac{1}{\gamma v_{0}}
\end{equation}
The curves meet smoothly at $X_{c}$. The curve $t'(X)$ for the first half of Alice's reckoning of Bill's proper time is given by
\begin{equation}\label{eq:Alice_Bill_out}
t' = \begin{cases}
\dfrac{c}{g}\sqrt{1 - \Bigl(1 + \dfrac{gX}{c^2}\Bigr)^{2}}, &X_{1} \le X \le 0\rule[-3.5ex]{0pt}{3.5ex}\\
t'_{1} - \dfrac{1}{\gamma v_{0}}(X - X_{1}), &X_{c} \le X \le X_{1}\rule[-3.5ex]{0pt}{3.5ex}\\
t'_{0} - \dfrac{c}{g}\sqrt{\Bigl(1 + \dfrac{gL}{c^{2}}\Bigr)^{2} - \Bigl(1 - \dfrac{gX}{c^{2}}\Bigr)^{2}}, &-L \le X < X_{c}
\end{cases}
\end{equation}
\noindent For the second half, replace $t'$ by $\Delta t_{\text{Bill}} - t'$, and solve for $t'$.

\end{appendices}

\end{document}